\documentclass[english,11pt,a4paper]{article}
\usepackage[T1]{fontenc}
\usepackage{jheppub}
\usepackage{babel}
\usepackage{amsmath}
\usepackage{amsfonts}
\usepackage{amstext}
\usepackage{amssymb}
\usepackage{mathrsfs}
\usepackage{diagbox}
\usepackage{tikz}
\usepackage{cancel}

\def\la{\langle}
\def\ra{\rangle}
\def\nn{\nonumber}
\def\tr{\text{tr}}
\def\Tr{\text{Tr}}
\newcommand{\qfac}[1]{[#1]_q!}
\newcommand{\qbinom}[2]{\begin{bmatrix}#1\\#2\end{bmatrix}_{\!q}}

\title{The Late-time Ramp of the Double-Scaled SYK Model from the large-$n$ tail of Cactus Diagram}

\author[a,b]{\large~ Yao Li}
\emailAdd{neolee@mit.edu}
\affiliation[a]{Center for Theoretical Physics, Massachusetts Institute of Technology,Cambridge, MA 02139, USA}
\affiliation[b]{INPAC, School of Physics and Astronomy, Shanghai Jiao Tong University, Shanghai 200240, China}

\abstract{
\\[1mm]
We derive the late-time ramp of the finite-temperature spectral form factor in the double-scaled SYK model directly from cactus diagrams, which is introduced as a multi-trace generalization of chord diagram in Ref \cite{Berkooz:2020fvm}. Although single-trace observables admit an exact chord-diagram description, multi-trace sums are obstructed by the chord-intersection weights $q_{IJ}$, which cannot be averaged independently to $q$. Our key idea is that the non-analytic contribution responsible for the ramp is controlled by the large-order tail of the Cactus-diagram expansion and is insensitive to finite changes in its low-order analytic terms. Decomposing the contribution with $n$-cross-trace-pairings Cactus diagram into two buds $B_n$ and a kernel $\mathcal K_n$, we determine their large-$n$ asymptotics with fixed $0\leq q<1$ and resum the resulting tail. For $\beta_L=\beta+it$ and $\beta_R=\beta-it$, we obtain
\begin{equation*}
Z_s^{\mathrm{sing}}(\beta+it,\beta-it)
=s_p c_N
\frac{|t|}{2\pi} \int_{E_{min}}^{E_{max}} e^{-2\beta E} dE+O(1),
\qquad
s_p=\begin{cases}2,&4\mid p,\\1,&4\nmid p.\end{cases}
\end{equation*}
This result reproduces the linear ramp predicted by random matrix theory, including its temperature dependence and symmetry factor, and agrees with the semiclassical predictions. It provides a direct microscopic origin of the ramp within the general $q$-deformed quantum algebra. While the plateau lies beyond the scope of the present analysis and calculation.
}

\keywords{\\[1mm]
double scaled SYK model, spectral form factor, late-time ramp, quantum chaos
}

\begin{document}
\maketitle
\section{Introduction}
The SYK model \cite{Sachdev:1992fk,Sachdev:2010um,Kitaev2015SimpleModel,Maldacena:2016hyu} provides a solvable, quantum-chaotic laboratory for low-dimensional holography. In the double-scaled limit, it is exactly tractable with chord diagrams. And the chord-diagram combinatorics naturally realizes a $q$-deformed quantum algebra.

Whereas the low-energy sector of conventional large-$p$ SYK is governed by Schwarzian quantum mechanics and is conjectured to be dual to JT gravity, double-scaled SYK contains a $q$-deformed Schwarzian theory and is conjectured to be dual to sine-dilaton gravity \cite{Blommaert:2024ymv}. Its auxiliary chord-number variable maps, in the low-energy limit, to a bulk geodesic-length variable \cite{Lin:2022rbf}. Both JT gravity and sine-dilaton gravity are expected to admit random-matrix descriptions, in which topological recursion \cite{Eynard2004Topological,EynardOrantin2007Invariants,EynardOrantin2009Topological,Eynard2014Invariants,EynardOrantin2007WeilPetersson} determines its multi-boundary amplitudes \cite{Saad:2019lba,Okuyama:2025hsd}.

The central observable in this comparison is the spectral form factor (SFF) \cite{Cotler:2016fpe}. For a Hamiltonian with energy levels $E_m$, the (unnormalized) finite-temperature SFF is
\begin{align}
\text{SFF}(\beta,t)
&\equiv \left\langle Z(\beta+it)Z(\beta-it)\right\rangle \nn \\
&=\left\langle\sum_{m,n}e^{-\beta(E_m+E_n)}e^{-it(E_m-E_n)}\right\rangle,
\qquad Z(\beta)=\tr e^{-\beta H}.
\end{align}
Here the bracket $\la \dots \ra$ includes the classical ensemble average and quantum average. It is the Fourier transform of the thermally weighted two-level spectral correlation function. The averaged density of states contains only one-level information, whereas the SFF probes correlations between pairs of energy levels and is therefore sensitive to level repulsion and long-range spectral rigidity. It is useful to separate the disconnected contribution, $\langle Z(\beta+it)\rangle\langle Z(\beta-it)\rangle$, from the connected part. The former is controlled by the smooth density of states, while the latter isolates genuine spectral correlations from spacetime wormholes.

The resulting time dependence provides a particularly sharp probe of quantum chaos. The spectral form factor should have the following universal structures: 
\begin{itemize}
\item  At early times, dephasing of the smooth density of states produces the decaying \textbf{slope}.
\item At intermediate times, universal two-level correlations generate the \textbf{ramp} predicted by random-matrix theory.
\item Near the Heisenberg time, when the evolution resolves individual level spacings, the ramp saturates into a \textbf{plateau} which is set by the discreteness of the spectrum.
\end{itemize}
Thus the slope--ramp--plateau structure makes the microscopic spectral information become visible. In holography, the ramp is  a diagnostic of a two-boundary observable whose connected part is associated semiclassically with wormhole or double-cone geometries \cite{Saad:2019lba,Khramtsov:2020bvs}. Deriving the ramp directly in a microscopic theory therefore provides a stringent test of how universal random-matrix correlations and nonperturbative bulk physics emerge from the same degrees of freedom. And the plateau probes the intrinsic discreteness of quantum gravity with continuous variables, which is naturally expected by the finite entropy and AdS/CFT correspondence.

The multi-trace correlators of double-scaled SYK, however, do not straightforwardly reproduce generic multi-boundary amplitudes \cite{Berkooz:2020fvm}. A naive chord-diagram expansion of the SFF reproduces the slope and its perturbative corrections but does not generate the ramp. By contrast, topological recursion yields the leading slope without these perturbative corrections and produces the ramp at the first perturbative order. This mismatch reflects the different organizations of the two expansions: the ramp is nonperturbative in SYK but perturbative in the topological expansion of random matrix as well as 2D dilaton gravity, whereas the plateau is nonperturbative in both descriptions. A path-integral formulation of double-scaled SYK provides a complementary semiclassical approach and reproduces the slope and ramp using distinct ans\"atzes \cite{Saad:2018bqo,Khramtsov:2020bvs}. It does not, however, explain how the ramp arises from the underlying $q$-deformed quantum mechanics or from the chord-diagram expansion.
And no one have successfully reproduced the plateau with path-integral formulation either. 

Besides, the ramp and plateau can also be derived from the smooth-filter framework of Ref.~\cite{Liu:2025ikq}. Current paper \cite{Liu:2026rpw} extracts the ramp and plateau of a single chaotic system as smooth macroscopic structures hidden in erratic microscopic spectral data. This viewpoint is complementary to the present ensemble-averaged calculation. Its derivation is valid for general chaotic systems, assuming the existence of the filter function.  Whereas our goal in this work is to identify the nonanalytic large-order structure that produces the ramp directly in the microscopic cactus-diagram expansion of double-scaled SYK.

Recent work \cite{Raz:2025wjw} also tried to provide a chord-diagram derivation to the late time ramp. The authors reduce the double-trace problem to a single-trace problem by introducing a complete operator basis $X_i$,
\begin{equation}
\tr(\mathcal O_1) \tr(\mathcal O_2)=\sum_i \tr(\mathcal O_1 X_i \mathcal O_2 X_i).
\end{equation}
The subsequent chord-diagram calculation relies on the absence of triple intersections, a condition that is difficult to establish for general $X_i$. The ramp is derived explicitly at $q=0$, while the result for general $q$ is obtained by generalizing the $q=0$ answer. This paper motivates us to seek a direct derivation from the cactus diagrams introduced in Ref.~\cite{Berkooz:2020fvm} for multi-trace correlators. We focus here on the ramp. Although the plateau can be understood heuristically as a consequence of the discrete spectrum of a finite-dimensional Hilbert space, its microscopic derivation likely requires the full set of finite-$N$ corrections and lies beyond the scope of this work.

In the cactus-diagram expansion, the spectral form factor takes the form
\begin{equation}
\text{SFF}=\sum_n C_n,
\end{equation}
where $C_n$ is the contribution with $n$ cross-trace pairings. At finite $n$, the intersection factors $q_{IJ}$ are correlated and therefore cannot be replaced independently by their average $q$, making a direct summation difficult. Our central idea is that the late-time ramp is encoded in the singular part of the resummed series. Finite changes to the low-order coefficients affect only analytic terms, so the singular part should be insensitive to a large cutoff $\Lambda$. This motivates the following replacement
\begin{equation}
\left.\text{SFF}\right|_{\mathrm{sing}}
\sim \left.\sum_{n>\Lambda}C_n\right|_{\mathrm{sing}}
\sim \left.\sum_{n\geq 0}C_n^{(\infty)}\right|_{\mathrm{sing}},
\end{equation}
where $C_n^{(\infty)}$ denotes the large-$n$ asymptotic form of $C_n$. Although this replacement does not determine the full spectral form factor, it retains the nonanalytic contribution that produces the nonperturbative ramp.

We implement this idea by factorizing $C_n$ into two buds $B_n$ and a kernel $\mathcal K_n$, deriving their large-$n$ asymptotics, and resumming the resulting singular contribution. The calculation yields a linear $|t|$ ramp with the expected temperature dependence and the symmetry factor determined by $p \mid 4$. This paper is organized as follows: Section~2 reviews double-scaled SYK, chord diagrams, etc. Section~3.1 introduces Cactus diagrams and the bud--kernel decomposition. Sections~3.2 and 3.3 derive the large-$n$ limits of the buds and kernels, respectively. Section~3.4 resums these asymptotic contributions and extracts the ramp. Section~4 summarizes the results and discusses the absence of plateau in this work.

\section{The double-scaled SYK model}
The SYK model is a quantum-mechanical system of $N$ Majorana fermions with random $p$-body couplings, where $p$ is assumed to be even. Its Hamiltonian is
\begin{equation}
	H=(i)^{\frac{p}{2}} \sum_I J_I \Psi_I.
\end{equation}
Here $I=i_1i_2\cdots i_p$ is a multi-index with $i_1<i_2<\cdots<i_p$, and
\begin{align}
	J_I&=j_{i_1 i_2 ... i_p}, \\
	\Psi_I &=\psi_{i_1}...\psi_{i_p}.
\end{align}
In early studies of the SYK model \cite{Sachdev:1992fk,Sachdev:2010um,Maldacena:2016hyu}, the couplings are drawn from a Gaussian ensemble with
\begin{align}
	\la J_I \ra_c &=0, \\
	\la J_I J_K \ra_c &=\mathcal J^2 \delta_{IK} =\frac{\bar{J}^2 \delta_{IK}}{\binom{N}{p}}=\frac{2\tilde{J}^2 p^2  \delta_{IK}}{N\binom{N}{p}}, 
\end{align}
where $\la\cdot\ra_c$ denotes the classical ensemble average. There are two normalization conventions are common in the literature. The parameter $\bar J=1$ is often used in chord-diagram calculations \cite{Berkooz:2018qkz,Berkooz:2018jqr,Berkooz:2020fvm}, whereas $\tilde J=1$ is convenient for extracting semiclassical bulk physics \cite{Lin:2022rbf}. We use the convention $\bar J=1$ throughout this paper.

The double-scaled limit is defined by
\begin{equation}
\label{eq:double-scaling}
p\to\infty,\qquad N\to\infty,\qquad \lambda=\frac{2p^2}{N}=\text{constant}.
\end{equation}
We will see that the SYK model is exactly tractable in this limit through chord diagrams.

\subsection{Chord diagrams}
We study the double-scaled SYK model through the finite temperature partition function
\begin{equation}
	Z_\beta=\langle \tr[e^{-\beta H}] \rangle,
\end{equation} 
where $\beta$ is the inverse temperature. Evaluating this partition function requires the moments of $H$:
\begin{align}
	m_k&=\langle H^k \rangle \nn \\
	&= i^{\frac{pk}{2}} \sum_{I_1,...,I_k} \langle J_{I_1}...J_{I_k}\rangle_c \langle \tr(\psi_{I_1}...\psi_{I_k})\rangle,k\in N.
\end{align}
The disorder average $\langle J_{I_1}\cdots J_{I_k}\rangle_c$ contributes a small factor $\binom{N}{p}^{-k/2}$, while summing the paired indices contributes $\binom{N}{p}^{[k/2]}$. Because the independent couplings have zero mean, configurations in which the $J_{I_i}$ occur in pairs dominate. For even $k$, these factors cancel; for odd $k$, a residual factor $\binom{N}{p}^{-1/2}$ vanishes in the double-scaled limit. We therefore restrict to even $k$, for which
\begin{align}
  m_k&= i^{\frac{pk}{2}}\mathcal{J}^k \binom{N}{p}^{-\frac{k}{2}} \sum_{I_1,...,I_k} \langle \tr(\psi_{I_1}...\psi_{I_k})\rangle,
\end{align}
where only $k/2$ of the $I_i$ are distinct.

The canonical anticommutation relation is
\begin{equation}
	\{\psi_i,\psi_j\}=2\delta_{ij}.
\end{equation}
This exact relation allows pure fermionic models, including double-scaled SYK, to be evaluated by Wick contractions. In particular,
\begin{equation}
	\langle \text{tr}(\psi_i \psi_j) \rangle =\delta_{ij}.
\end{equation} 
Consider first the simple case $\la\tr(\psi_I\psi_I)\ra$, with no sum over $I=i_1\cdots i_p$. Reordering the fermions introduces signs, and a careful count gives
\begin{equation}
	\la \tr(\psi_I \psi_I) \ra = (-1)^{\frac{p^2-p}{2}}=(-1)^{-\frac{p}{2}}.
\end{equation}
In the last equality we used the fact that $p$ is even. The same argument applies whenever identical $I_i$ are adjacent, yielding
\begin{align}
	&i^{\frac{kp}{2}} \la \tr(\psi_{I_1} \psi_{I_1} \psi_{I_2} \psi_{I_2} ... \psi_{I_{k/2}} \psi_{I_{k/2}}) \ra \nn \\
	=& (-1)^{\frac{kp}{4}} \times (-1)^{-\frac{p}{2}\times \frac{k}{2}} \nn \\
	=& 1.
\end{align}

We therefore evaluate a general expression $\la\tr(\psi_{I_1}\psi_{I_2}\cdots)\ra$ by reordering it into the paired form $\la\tr(\psi_{I_1}\psi_{I_1}\cdots\psi_{I_{k/2}}\psi_{I_{k/2}})\ra$. The sign acquired when exchanging $\psi_I$ and $\psi_J$ depends on the overlap $m=|I\cap J|$:
\begin{equation}
	\psi_{I}\psi_{J}  =  (-1)^m\psi_J \psi_I.
\end{equation}  
Averaging over $I$ and $J$, the probability of a given $m$ is
\begin{equation}
	p(m)=\frac{\binom{N}{p} \binom{p}{m} \binom{N-p}{p-m}}{\binom{N}{p}^2}\simeq \frac{(\lambda/2)^m}{m!}e^{-\lambda/2},
\end{equation}
where $m\ll p\ll N$ and $2p^2/N=\lambda$ is held fixed. It then follows that
\begin{equation}
	\la (-1)^m \ra_c=\sum_m \frac{(\lambda/2)^m}{m} e^{-\lambda/2} (-1)^m=e^{-\lambda} \equiv q.
\end{equation}
We define $q=e^{-\lambda}$ for later convenience.

A general expression $\la\tr(\psi_{I_1}\psi_{I_2}\cdots)\ra$ is conventionally represented by a chord diagram. For example,
\begin{align}
	&\begin{tikzpicture}[xscale=1,yscale=1]
		\draw [thick] (0,0) circle [radius=2];
		\filldraw [thick] (0,2) circle [radius=2pt]
		(0,-2) circle [radius=2pt]
		(2,0) circle [radius=2pt]
		(-2,0) circle [radius=2pt];
		\draw (-2,0) .. controls (-1,0) and (0,-1) .. (0,-2);
		\draw (2,0) .. controls (1,0) and (0,1) .. (0,2);
		\draw (-2.2,0) node {$1$};
		\draw (0,-2.3) node {$2$};
		\draw (2.2,0) node {$3$};
		\draw (0,2.3) node {$4$};
		\draw (-4.5,0) node {$\la \tr(\psi_{I_1} \psi_{I_1} \psi_{I_2} \psi_{I_2})\ra=$};
	\end{tikzpicture}, \\
	&\begin{tikzpicture}[xscale=1,yscale=1]
		\draw [thick] (0,0) circle [radius=2];
		\filldraw [thick] (0,2) circle [radius=2pt]
		(0,-2) circle [radius=2pt]
		(2,0) circle [radius=2pt]
		(-2,0) circle [radius=2pt];
		\draw (-2,0) --(2,0);
		\draw (0,2) -- (0,-2);
		\draw (-2.2,0) node {$1$};
		\draw (0,-2.3) node {$2$};
		\draw (2.2,0) node {$3$};
		\draw (0,2.3) node {$4$};
		\draw (-4.5,0) node {$\la \tr(\psi_{I_1} \psi_{I_2} \psi_{I_1} \psi_{I_2})\ra=$};
	\end{tikzpicture}.
\end{align} 
The circle represents the trace, and each $\psi_{I_i}$ is represented by a node labeled $i$. If $I_i=I_j$, a chord connects nodes $i$ and $j$. The number of chord intersections equals the number of exchanges required to reach the standard diagram in which every chord connects adjacent nodes. Consequently,
\begin{equation}\label{eq:mk-diagram}
	m_k=\sum_{\text{chord diagrams}} q^{\text{number of intersections}}.
\end{equation} 
Equation~\eqref{eq:mk-diagram} is simple and elegant, but it does not by itself provide an analytic solution. An algebraic reformulation of the chord-diagram sum does, and we review that construction in the next section.

We may also introduce an observable $M_A$ built from polynomials in the $\psi_i$:
\begin{equation}
	M_A=i^{\frac{p_A}{2}}\sum_{i_1,...,i_{p_A}} J^A_{i_1,...,i_{p_A}} \psi_{i_1}...\psi_{i_{p_A}},
\end{equation}
where $J^A$ is drawn from an ensemble analogous to that of $J$, so that $M_A$ couples consistently to $H$. We require
\begin{align}
	\la J^A_I \ra_c &=0, \\
	\la J^A_I J^A_I \ra_c &= \binom{N}{p_A}^{-1} \bar{J}^2_A,
\end{align}
and 
\begin{equation}
	\lambda_A=\frac{2p^2_A}{N}=\text{constant}.
\end{equation}
Additional observables $M_B,M_C,\ldots$ can be introduced in the same way.

The general multipoint correlation function is
\begin{align}
	\mathcal  A_n&=\la \tr(e^{-\beta H} M_1(t_1) M_2(t_2)...M_n(t_n)) \ra , \\
	M_i(t_i)&= e^{it_i H} M_i e^{-it_i H}.
\end{align}
Expanding $e^{-\beta H}$ or $e^{itH}$ gives the chord-diagram representation
\begin{align}\label{eq:An-diagram}
	&\la \tr(H^{k_1} M_1 H^{k_2} M_2 H^{k_3}...) \ra \nn \\
	=&\sum_{\text{chord diagrams}} \prod_{i,j\geq 0} q^{\text{No. of i-j chord intersections}}. 
\end{align}
We treat the $M_i$ as additional nodes in the chord diagram. Two identical observables with the same index sequence are connected by a new type of chord, called an $i$-chord; the original Hamiltonian chords are called $0$-chords. For example,
\begin{equation}
	\begin{tikzpicture}[xscale=1,yscale=1]
		\draw [thick] (0,0) circle [radius=2];
		\filldraw [thick] (0,2) circle [radius=2pt]
		(0,-2) circle [radius=2pt]
		(2,0) circle [radius=2pt]
		(-2,0) circle [radius=2pt];
		\filldraw [red]
		(-1.4142,-1.4142) circle [radius=2pt]
		(1.4142,1.4142) circle [radius=2pt];
		\draw (-2,0) .. controls (-1,0) and (0,-1) .. (0,-2);
		\draw (2,0) .. controls (1,0) and (0,1) .. (0,2);
		\draw [dash dot,red] (1.4142,1.4142) -- (-1.4142,-1.4142);
		\draw (-2.2,0) node {$1$};
		\draw (0,-2.3) node {$2$};
		\draw (2.2,0) node {$3$};
		\draw (0,2.3) node {$4$};
		\draw (-1.58,-1.58) node {$M_1$};
		\draw (1.58,1.58) node {$M_1$};
		\draw (-5,0) node {$\la \tr(\psi_{I_1}M_1 \psi_{I_1} \psi_{I_2}M_1 \psi_{I_2})\ra=$};
	\end{tikzpicture}.
\end{equation}

\subsection{Transfer matrix}
Equations~\eqref{eq:mk-diagram} and \eqref{eq:An-diagram} are simple and elegant, but individual chord diagrams carry little analytic information. The physically relevant object is the sum over all diagrams with specified numbers of $H$ and $M_i$ insertions. Remarkably, this full sum is analytically tractable even when a generic individual diagram is not.

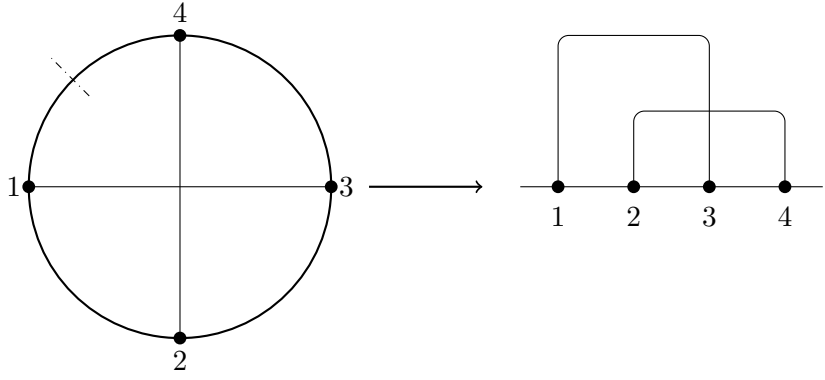
\begin{figure}
	\centering
	\begin{tikzpicture}[xscale=1,yscale=1]
		\draw [thick] (0,0) circle [radius=2];
		\filldraw [thick] (0,2) circle [radius=2pt]
		(0,-2) circle [radius=2pt]
		(2,0) circle [radius=2pt]
		(-2,0) circle [radius=2pt];
		\draw (-2,0) --(2,0);
		\draw (0,2) -- (0,-2);
		\draw (-2.2,0) node {$1$};
		\draw (0,-2.3) node {$2$};
		\draw (2.2,0) node {$3$};
		\draw (0,2.3) node {$4$};
		\draw [dash dot](-1.2,1.2) -- (-1.7,1.7);
		\draw [thick,->] (2.5,0) -- (4,0);
		\draw (4.5,0)-- (8.5,0);
		\draw (5,-0.4) node {$1$};
		\draw (6,-0.4) node {$2$};
		\draw (7,-0.4) node {$3$};
		\draw (8,-0.4) node {$4$};
		\filldraw [thick] (5,0) circle [radius=2pt]
		(6,0) circle [radius=2pt]
		(7,0) circle [radius=2pt]
		(8,0) circle [radius=2pt];
		\draw [rounded corners] (5,0) -- (5,2) -- (7,2) -- (7,0);
		\draw [rounded corners] (6,0) -- (6,1) -- (8,1) -- (8,0);
	\end{tikzpicture}
	\caption{The dashed line indicates where the circular chord diagram is cut. \emph{We adopt the standard convention that a chord beginning at node $i$ cannot intersect chords already above node $i$.}}
	\label{fig:cut-chord}	
\end{figure}

We first consider chord diagrams without observable insertions. Cut the circular diagram at a reference point, as shown in Fig.~\ref{fig:cut-chord}, and regard the resulting diagram as a dynamical system evolving from left to right. We call the segment between the cut and node $1$ stage $0$, the segment between nodes $i$ and $i+1$ stage $i$, and the segment between node $k$ and the cut stage $k$. Let $v_l^{(i)}$ denote the sum of all diagrams from stage $0$ to stage $i$ with $l$ chords above stage $i$. Then
\begin{equation}
	v_{l}^{(0)}=\delta_{l,0}.
\end{equation} 
This equation supplies the initial condition for the recursion relation satisfied by $v_l^{(i)}$.

To derive the recursion, move from stage $i$ to stage $i+1$, with $i\neq0,k-1$. There are only two possibilities:
\begin{enumerate}
	\item A chord starts at node $i+1$;
	\item A chord ends at node $i+1$.
\end{enumerate}
Suppose that there are $l$ chords above stage $i+1$. In the first case, our convention introduces no new intersections, and there are $l-1$ chords above stage $i$. In the second case, there are $l+1$ chords above stage $i$, and the number of new intersections ranges from $0$ to $l$. Hence $v_l^{(i)}$ satisfies
\begin{equation}\label{eq:recursive-v1}
	v_l^{(i+1)}=v^{(i)}_{l-1}+(\sum_{n=0}^l q^n) v^{(i)}_{l+1}=v^{(i)}_{l-1}+\frac{1-q^{l+1}}{1-q} v^{(i)}_{l+1},v^{(i)}_{l<0}=0.
\end{equation}
Therefore, $m_k=v_0^{(k)}$.

Define an auxiliary Hilbert space spanned by states $|l\ra$, where $0\leq l\leq k/2$ denotes the number of chords above the stage under consideration. There are $s=k/2+1$ such states. Equation~\eqref{eq:recursive-v1} can then be interpreted as matrix multiplication:
\begin{equation}
	v^{(i+1)}=T_{s \times s} v^{(i)},
\end{equation}  
where $v^{(i)}$ is expanded in the basis $|l\ra$, and the matrix $T_{s\times s}$ is
\begin{equation}
	T_{s \times s }=\begin{bmatrix}
		0 & \frac{1-q}{1-q} & 0 & 0 &  \\
		1 & 0 & \frac{1-q^2}{1-q} & 0& ... \\
		0 & 1 & 0 & \frac{1-q^3}{1-q} &  \\
		&   & ... &  &  
	\end{bmatrix}.
\end{equation}
The moment $m_k$ can therefore be written as
\begin{equation}
	m_k= \la 0 | T^k_{s \times s } | 0 \ra.
\end{equation}
\subsection{Eigenvectors and eigenvalues}
Although $T_{s\times s}$ is the smallest matrix required to evaluate $m_k$, any $T_{n\times n}$ with $n>k/2$ gives the same result. It is therefore sufficient to study $T=T_{\infty\times\infty}$, for which
\begin{equation}
	m_k= \la 0 | T^k | 0 \ra.
\end{equation}   

The advantage of the infinite-dimensional matrix is that discrete spectral sums become integrals when the eigenvalue spectrum becomes continuous as $k\to\infty$. Let an eigenvector of $T_{s\times s}$ be
\begin{equation}
	|\phi_E \ra =\sum_{l=0}^{k/2} A(l)|l\ra,
\end{equation}  	
with eigenvalue $E$. The equation $T_{s \times s}|\phi_E\ra=E|\phi_E\ra$ yields
\begin{equation}\label{eq:eigenvector}
	A(l)+\frac{1-q^{l-1}}{1-q}A(l-2)= E A(l-1).
\end{equation} 
Because an eigenvector is defined only up to an overall nonzero factor, we may set a chosen component $A(l_{\rm ref})$ to one. For eigenvectors with $A(0)\neq0$, Eq.~\eqref{eq:eigenvector} is supplemented by the initial conditions $A(0)=1$ and $A(1)=E$. Components with $l\geq s$ lie outside the finite-dimensional space, so one must also impose the terminal condition $A(s)=0$, equivalently $\frac{1-q^{s-1}}{1-q}A(s-2)=EA(s-1)$.
For instance, when $k=2$,
\begin{equation}
	T_{2\times 2}=\begin{bmatrix}
		0 & \frac{1-q}{1-q}  \\ 1 & 0 
	\end{bmatrix}
\end{equation} 
with eigenvalues $\pm1$, precisely as required by the constraint $A(2)=E^2-1=0$. This finite-dimensional approach, however, does not make the large-$k$ limit transparent.

Since $q=e^{-\lambda}<1$, the large-$k$ tail of $T_{\infty\times\infty}$ approaches
\begin{equation}
	T_{\infty \times \infty }=\begin{bmatrix}
		&...  & ...  &...  &... \\
		&...  &0   &\frac{1}{1-q}  &0 \\
		&...  &1  &0  &\frac{1}{1-q} \\
		&...  &0   &1  &0
	\end{bmatrix}
\end{equation}
For finite $k$, one may exchange the roles of the initial and terminal conditions and set $A(s-1)=1$. The same reasoning is expected to hold as $k\to\infty$.

Define the asymptotic matrix associated with $T_{\infty\times\infty}$ by
\begin{equation}
	T_{\infty \times \infty}^{\text{a}}=\begin{bmatrix}
		0 & \frac{1}{1-q} & 0 & 0 &  \\
		1 & 0 & \frac{1}{1-q} & 0& ... \\
		0 & 1 & 0 & \frac{1}{1-q} &  \\
		&   & ... &  &  
	\end{bmatrix}
\end{equation}
We argue that this matrix has the same eigenvalue spectrum as $T_{\infty\times\infty}$. In this limit, Eq.~\eqref{eq:eigenvector} becomes
\begin{equation}\label{eq:eigenvector2}
	A(l)+\frac{1}{1-q}A(l-2)= E A(l-1).
\end{equation}
Substituting the ansatz $A(l)\sim[r/\sqrt{1-q}]^l$, with $r\in\mathbb R$, gives
\begin{equation}
	r^2-2c\times r+1=0,2c=E\sqrt{1-q}.
\end{equation}
There are two solutions, $r_\pm=c\pm\sqrt{c^2-1}$, satisfying $r_+r_-=1$. Therefore,
\begin{equation}
	A(l)=C_+ r_+^l + C_- r_-^l=C_+ r_+^l + C_- r_+^{-l},
\end{equation}
where the constants $C_\pm$ are fixed by the initial and terminal conditions. The initial condition $A(-1)=0$ further implies $C_+/r_+ + C_-r_+=0$. For convenience, define
\begin{align}
	C'_+&=C_+/r_+ \\
	C'_-&=C_-*r_+=C_-/r_- 
\end{align}
As a consequence,
\begin{equation}
	A(l)=C'_+ r_+^{l+1} + C'_- r_-^{l+1}=C'_+ (r_+^l -  r_+^{-l-1})
\end{equation}

Now regulate the asymptotic matrix to $T^a_{s\times s}$. Its terminal condition is
\begin{equation}
	A(s)=C'_+ (r_+^{s+1} - r_+^{-s-1})=0.
\end{equation}
Writing $r_+=e^{i\theta}$, this condition becomes
\begin{equation}
	(r_+^{s+1} - r_+^{-s-1})=2i\sin((s+1)\theta)=0.
\end{equation}
The solutions are $\theta_n=n\pi/(s+1)$, with $n\in\mathbb N$.

The corresponding eigenvalues are
\begin{equation}
	E(\theta_n)= 2\frac{\cos(\theta_n)}{\sqrt{1-q}}
\end{equation}
As $s\to\infty$, the eigenvalues become dense in the interval
\begin{equation}
	E\in (-\frac{2}{\sqrt{1-q}},\frac{2}{\sqrt{1-q}})
\end{equation}
and the spectrum becomes continuous.

The matrices $T^a_{\infty\times\infty}$ and $T_{\infty\times\infty}$ do not have the same eigenvectors because their small-$l$ components differ. Their common eigenvalue spectrum nevertheless allows the parametrization
\begin{equation}
	E(\mu)=\frac{2\mu}{\sqrt{1-q}},\mu \in (-1,1)
\end{equation}
and Eq.~\eqref{eq:eigenvector} becomes
\begin{equation}
	A(l,\mu)+\frac{1-q^{l-1}}{1-q}A(l-2,\mu)= \frac{2\mu}{\sqrt{1-q}} A(l-1,\mu).
\end{equation}
Define
\begin{equation}
	A(l,\mu)=\frac{(1-q)^{\frac{l}{2}}}{(q;q)_l}B(l,\mu)
\end{equation}
where 
\begin{equation}
	(a;q)_l=\prod_{k=1}^{l}(1-aq^{l-1}).
\end{equation}
The function $B(l,\mu)$ satisfies the recursion relation for the continuous $q$-Hermite polynomials. Thus,
\begin{align}\label{eq:eigensystem-T}
	A(l,\mu)&=\frac{(1-q)^{\frac{l}{2}}}{(q;q)_l}H_l(\mu|q),\\
	|\phi(\mu)\ra &=\sum_l \frac{(1-q)^{\frac{l}{2}}}{(q;q)_l}H_l(\mu|q)|l\ra. 
\end{align}

We can now express products of infinite-dimensional matrices and vectors as spectral integrals:
\begin{align}
	m_k &= \la 0 | T^k |0 \ra \\
	&=\int d\mu \phi^2_0(\mu) E^k(\mu)
\end{align}
where $\phi_0(\mu)=\la 0 | \phi(\mu) \ra$.

The eigenvectors in Eq.~\eqref{eq:eigensystem-T} are not in the normalization most commonly used in the literature. Berkooz et al.~\cite{Berkooz:2018qkz} instead define
\begin{equation}
	\hat{T}=P.T.P^{-1},
\end{equation}
where the diagonal matrix $P$ is
\begin{equation}
	P_{ll}=\frac{\sqrt{(q;q)_l}}{(1-q)^\frac{l}{2}},l\neq 0
\end{equation}
with $P_{00}=1$. Hence
\begin{equation}
	m_k=\la 0|T^k|0\ra=\la 0|\hat T^k|0\ra.
\end{equation}
Since $P_{00}=1$, $|0\ra=|\hat0\ra$. In analogy with
\begin{equation}
	T=a+\hat{a},
\end{equation}
it is conventional to define 
\begin{equation}
	\hat{T}=A_+ + A_-.
\end{equation}

The eigenvectors of $\hat T$ are $|\hat\phi\ra=P|\phi\ra$ and have the same eigenvalues $E(\mu)$. After normalization,
\begin{equation}
	|\hat \phi_0|^2 =\frac{1}{2\pi}(q;q)_\infty |(e^{2i\theta};q)_\infty|^2=\frac{1}{2\pi}(q,e^{\pm 2i\theta};q)).
\end{equation}
Defining $d\mu(\theta)=|\hat\phi_0|^2d\theta$, we obtain
\begin{equation}
	m_k=\int d\mu(\theta) E(\theta)^k.
\end{equation}

\subsection{Skeleton diagram}
Although no general method is known for computing arbitrary $2n$-point functions, Berkooz et al.~\cite{Berkooz:2018jqr} extracted a set of diagrammatic rules for a special class of such correlators from explicit two- and four-point calculations. In literature, this construction is referred as the skeleton-diagram method.

There are two main differences between skeleton diagrams and ordinary chord diagrams. First, rather than using the chord-number basis $|l\ra$, one works in the energy basis $|\phi_\theta\ra$, with eigenvalue $E=2\cos\theta/\sqrt{1-q}$ and $q=q_{00}$. Hamiltonian chords are then omitted, and each segment is assigned a propagator:
\begin{equation}
	\begin{tikzpicture}
		\draw (0,0) arc [start angle=135,end angle=45,radius=2];
		\draw (1.3,0.8) node {$\theta$};
		\draw (0,-0.3) node {$\tau_1$};
		\draw (3,-0.3) node {$\tau_2$};
	\end{tikzpicture}
	=d\mu(\theta) e^{E(\theta)\Delta \tau },
\end{equation}
where 
\begin{equation}
	d\mu(\theta) = \frac{d\theta}{2\pi} (q,e^{\pm 2i\theta};q)_\infty.
\end{equation}

Second, one introduces an interaction vertex between adjacent segments and an $M$-chord:
\begin{equation}
	\begin{tikzpicture}
		\draw (0,0) arc [start angle=135,end angle=45,radius=2];
		\draw (0.8,0.8) node {$\theta_1$};
		\draw (1.8,0.8) node {$\theta_2$};
		\draw [dash dot,red] (1.3,-0.5)-- (1.3,0.7);
		\filldraw [red] (1.3,0.6) circle [radius=2pt];
	\end{tikzpicture}
	= \sqrt{\frac{(\tilde{q}^2;q)_\infty}{(\tilde{q}e^{\pm i\theta_1 \pm i\theta_2};q)_\infty}},
\end{equation}
where $\tilde{q}=q_{0,1}$. A crossing between internal $M$-chords requires an additional interaction vertex:
\begin{equation}
	\begin{tikzpicture}
		\draw (0,0) circle [radius=2];
		\draw [red,dash dot] (-1.414,1.414) -- (1.414,-1.414);
		\draw [red,dash dot] (1.414,1.414) -- (-1.414,-1.414);
		\draw (0,2.2) node {$\theta_1$};
		\draw (2.2,0) node {$\theta_2$};
		\draw (0,-2.2) node {$\theta_3$};
		\draw (-2.2,0) node {$\theta_4$};
		\draw (1.42,1.42) node {$\tau_1$};
		\draw (1.42,-1.42) node {$\tau_2$};
		\draw (-1.42,-1.42) node {$\tau_3$};
		\draw (-1.42,1.42) node {$\tau_4$};
		\draw (1.1,0.7) node {$\tilde{q}$};
		\draw (1.1,-0.7) node {$\bar{q}$};
	\end{tikzpicture}
	=R_{\theta_4 \theta_2}^{(2)}\begin{bmatrix}
		\theta_3 & l_1 \\
		\theta_1 & l_2
	\end{bmatrix}.	
\end{equation}
where $\tilde q=q^{l_1}$ and $\bar q=q^{l_2}$. The explicit expression for $R$ is lengthy and can be found in Ref.~\cite{Berkooz:2018jqr}. 

Finally, we remark that the Skeleton diagram method is only valid for a special set of chord diagrams, especially the chord diagrams without intersection of matter chords. Fortunately, only the chord diagrams without intersection of matter chords are required in this work.

\subsection{Reformulating chord algebra}
To generalize the chord space and algebra to multi-trace cases in later section, we reformulate the chord space and algebra as follows.

The random coupling $J_I$ can be represented by the operator-valued coupling
\begin{equation}
	\hat J_I = \mathcal J a_I + \mathcal J a^\dagger_I,
\end{equation}
where $a_I$ and $a_I^\dagger$ are annihilation and creation operators,
\begin{equation}\label{eq:commutation}
	[a_I,a_J^\dagger]=\delta_{IJ}.
\end{equation}
Define a quantum vacuum $|0\ra$ by
\begin{equation}
	\forall I, a_I|0\ra = \la 0| a^\dagger_I =0.	
\end{equation} 
It is then immediate that
\begin{align}
	\la \hat J_I \ra_J &=0, \\
	\la \hat J_I^2 \ra_J &= \mathcal{J}^2. 
\end{align}
Here $\la\cdot\ra_J$ denotes the expectation value in $\mathcal H_J$. This operator-valued coupling is simply a mathematically equivalent reformulation of the conventional random-coupling ensemble.

For later convenience, we define new operators
\begin{align}
	A &=\mathcal J \sum_I A_I = i^{p/2}\mathcal J \sum_I a_I \Psi_I, \\
	A^\dagger &=\mathcal J \sum_I A_I^\dagger = i^{p/2} \mathcal J \sum_I a^\dagger_I \Psi_I, \\
	\mathbb I &=i^{p} \mathcal J^2 \sum_I \Psi_I \Psi_I,
\end{align}
Here we use the operator identity $\mathbb I=i^p\mathcal J^2\Psi_I\Psi_I=1$. More generally,
\begin{equation}
	A_I A^\dagger_J=q_{IJ} A^\dagger_J A_I +\delta_{IJ}.
\end{equation}
Although $i^p\mathcal J^2\sum_{I,J}q_{IJ}a_J^\dagger\Psi_Ja_I\Psi_I\neq qA^\dagger A$ as an operator identity, we argue that in the double-scaled large-$N$ limit, vacuum expectation values of operators built from $\{A,A^\dagger,\mathbb I\}$ obey the effective relation
\begin{equation}\label{eq:qalgebra}
	AA^\dagger \rightarrow qA^\dagger A + \mathbb I.
\end{equation}
This relation follows because
\begin{enumerate}
	\item Once all $A_I=a_I\Psi_I$ and $A_I^\dagger=a_I^\dagger\Psi_I$ are contracted, the ensemble average involves only products and sums of $q_{IJ}$.
	\item In the double-scaled large-$N$ limit, triple intersections can be neglected, so distinct $q_{IJ}$ may be treated as independent variables and averaged to $q$.
\end{enumerate}
Equation~\eqref{eq:qalgebra} therefore holds only within the relevant expectation values, so we use $\rightarrow$ rather than $=$.

Remarkably, Eq.~\eqref{eq:qalgebra} coincides with the $q$-deformed chord algebra obtained from chord diagrams in double-scaled SYK \cite{Berkooz:2018qkz,Berkooz:2018jqr,Berkooz:2024lgq}. There is nevertheless a subtle distinction: the partition function is defined as a trace over $\mathcal H_\psi$, whereas the effective chord algebra represents it as a vacuum expectation value of the transfer matrix $T$:
\begin{equation}
	\la \tr e^{-\beta H}\ra_J=\la 0| e^{-\beta T} | 0\ra_c.
\end{equation} 
More precisely, both sides can be represented in the common Hilbert space $\mathcal H_J\otimes\mathcal H_\psi$:
\begin{equation}
	\la \tr e^{-\beta H}\ra_J=\la 0| e^{-\beta (A+A^\dagger)} | 0\ra_{J,\psi}.
\end{equation}

To establish this identity, represent the Majorana fermions as tensor products of Pauli matrices:
\begin{align}
	\psi_1 &= \sigma^1\otimes \sigma^0 \otimes ... \otimes \sigma^0, \\
	\psi_2 &= \sigma^3\otimes \sigma^1 \otimes ... \otimes \sigma^0, \\
	& \dots \nn \\
	\psi_N &= \sigma^3\otimes \sigma^3 \otimes ... \otimes \sigma^1.
\end{align}
Every nontrivial operator is a linear combination of the basis monomials
\begin{align}
	\mathcal O_P=\psi_{i_1}...\psi_{i_p},i_1<i_2<...<i_p\in \{1,...,N\}.	
\end{align}
Here $P=\{i_1,\ldots,i_p\}$. For even $p$,
\begin{equation}
	\mathcal O_P=\prod_j^N \otimes \sigma^{s_j},\sigma^{s_j}=\begin{cases}
		i\sigma^2,s_j\in P, \\\sigma^0,s_j \notin P.
	\end{cases}
\end{equation}
The matrices $i\sigma^2$ and $\sigma^0$ satisfy
\begin{align}
	&\frac{1}{2}\Tr(i\sigma^2)=\la 0|i\sigma^2 |0\ra=\la 1|i\sigma^2 |1\ra=0, \\
	&\frac{1}{2}\Tr(\sigma^0)=\la 0|\sigma^0 |0\ra=\la 1|\sigma^0 |1\ra=1
\end{align}
It follows that
\begin{equation}
	\tr \mathcal O_P=\frac{1}{2^N}\Tr \mathcal O_P=\la \Omega| \mathcal{O}_P |\Omega\ra=\begin{cases}
		1,p=0, \\ 0,p\neq 0.
	\end{cases}
\end{equation}
where $|\Omega\ra=\prod_i^N\otimes|0\ra_i$. For odd $p$,
\begin{equation}
	\mathcal O_P=\prod_j^N \otimes \sigma^{s_j},\sigma^{s_j}=\begin{cases}
		\sigma^1,s_j\in P, \\\sigma^0\text{ or }\sigma^3,s_j \notin P.
	\end{cases}
\end{equation}
The presence of $\sigma^1$ implies
\begin{equation}
	\tr  \mathcal O_P=\la \Omega| \mathcal O_p |\Omega\ra=0.
\end{equation}
Thus, for every $\mathcal O_P$,
\begin{equation}
	\tr  \mathcal O_P=\la \Omega| \mathcal O_p |\Omega\ra.
\end{equation}
The state $|\Omega\ra$ may in fact be replaced by any computational-basis state $|s\ra$:
\begin{equation}
	|s\ra = \prod_i^N \otimes |s_i\ra_i, s_i=0,1.
\end{equation}
These identities between traces and expectation values are a consequence of particle-hole duality.

\section{The double-trace correlation function}
\subsection{Cactus diagrams}
The operators satisfy
\begin{align}
	A_L A^\dagger_R &= A^\dagger_R A_L + \mathbb I_{LR}, \\
	A_R A^\dagger_L &= A^\dagger_L A_R + \mathbb I_{LR},	
\end{align}  
because $\Psi_I^L$ and $\Psi_J^R$ commute. Here we define
\begin{equation}
	\mathbb{I}_{LR}=\mathbb J^2 \sum_I \mathbb I_{LR}^I=i^{p} \mathcal J^2\Psi_I^L\Psi_I^R.	
\end{equation}

Unlike $\mathbb I=\mathbb I_{LL}=\mathbb I_{RR}=1$, the operator $\mathbb I_{LR}$ is nontrivial and encodes the wormhole interaction between distinct traces. It also satisfies
\begin{align}\label{eq:LRchord-1}
	A^I_L \mathbb I^J_{LR}&=q_{IJ} \mathbb I^J_{LR} A^I_L, \\
	\label{eq:LRchord-2}
	A^I_R \mathbb I^J_{LR}&=q_{IJ} \mathbb I^J_{LR} A^I_R, \\
	\label{eq:LRchord-3}
	\mathbb I^J_{LR} (A^\dagger_L)^I &= q_{IJ} (A^\dagger_L)^I \mathbb I^J_{LR}, \\
	\label{eq:LRchord-4}
	\mathbb I^J_{LR} (A^\dagger_R)^I &= q _{IJ}(A^\dagger_R)^I \mathbb I^J_{LR}.
\end{align}
These relations allow the multi-trace correlation function be expressed by similar chord diagrams. But the significant difference is that  $q_{IJ}$ cannot be averaged independently when $\mathbb I_{LR}^I$ is present.

For instance,
\begin{align}
	\label{eq:K2}
	&\la 0|\tr(A_L A_L A_R^\dagger A_R^\dagger) |0\ra  \nn \\	
	=&\mathcal{J}^4 \sum_{I_1,I_2,I_3,I_4} \la 0| \tr (A_L^{I_1}A_L^{I_2} (A_R^\dagger)^{I_3}(A_R^\dagger)^{I_4}) |0\ra \nn \\
	=&\mathcal{J}^4 \sum_{I_1,I_2,I_3,I_4} \tr(\mathbb I_{LR}^{I_1} \mathbb I_{LR}^{I_2} ) (\delta_{I_1 I_3}\delta_{I_2 I_4} + \delta_{I_2 I_3}\delta_{I_1 I_4}q_{I_1 I_2}) \nn \\
	\neq & \tr(\mathbb I_{LR} \mathbb I_{LR} ) (1+q).
\end{align}
Because $\mathcal H_\psi=\mathcal H_{\psi^L}\otimes\mathcal H_{\psi^R}$, we use $\tr=\tr_L\otimes\tr_R$. In this example, $\tr(\mathbb I_{LR}^{I_1}\mathbb I_{LR}^{I_2})=\delta_{I_1I_2}$ can be evaluated explicitly, and $q_{I_1I_2}=1$. Hence
\begin{align}
	\la 0|\tr(A_L A_L A_R^\dagger A_R^\dagger) |0\ra =	2 \mathcal J^2.
\end{align} 
The two $\mathbb I_{LR}^I$ chords connecting the left and right traces define a $2$-pairing. A $1$-pairing vanishes because it is proportional to
\begin{equation}
	\tr \mathbb I_{LR}=0.	
\end{equation}
Thus $2$-pairings provide the leading contribution to the connected part of the double-cone partition function. In general,
\begin{equation}
	\la 0| \tr(H_L^{k_L}H_R^{k_R}) |0\ra|_{2-\text{pairings}}=\mathcal{J}^2\frac{k_L k_R}{2} \la 0| \tr(H_L^{k_L}) |0\ra \la 0|\tr(H_R^{k_R}) |0\ra
\end{equation}
where $\frac{k_L}{2}\times\frac{k_R}{2}$ counts the possible choices of the $2$-pairing. Therefore,
\begin{equation}
	Z(\beta_L,\beta_R)|_{2-\text{pairings}}=\frac{\mathcal{J}^2}{2} \beta_L\frac{\partial}{\partial \beta_L} Z(\beta_L) \beta_R                          \frac{\partial}{\partial \beta_R} Z(\beta_R)
\end{equation}
This result was firstly obtained in Ref.~\cite{Berkooz:2020fvm}, which also introduced the graphical representation known as Cactus diagrams. In our convention, these Cactus diagrams can be identified with
\begin{equation}
	\la m| \tr(\hat Z) |n\ra=	
	\begin{tikzpicture}{xscale=1;yscale=1}
		\filldraw (-0.15,-0.15) rectangle (0.2,0.2);
		\draw (0,0) -- (1,1) circle [radius=0.1];
		\draw (0,0) -- (1,-1) circle [radius=0.1];
		\filldraw (0,0) -- (-1,1) circle [radius=0.1];
		\filldraw (0,0) -- (-1,-1) circle [radius=0.1];
		\draw (-1.5,0) node {$m \Big \{ \vdots$};
		\draw (1.5,0) node {$\vdots$ \Big \} n};
	\end{tikzpicture}.
\end{equation} 
Here we distinguish creation nodes from annihilation nodes.

The double-trace correlation function can be written as
\begin{align}
Z(\beta_L,\beta_R)=&
\sum_{n\geq 2} Z(\beta_L,\beta_R)_{n-\text{pairings}} \nn \\
=&\la \tr (e^{-\beta_L (A_L +A_L^\dagger)} e^{-\beta_R(A_R +A_R^\dagger)}) \ra \nn \\
=&\sum_{n\geq 2} B_n(\beta_L) \mathcal K_n B_n(\beta_R).
\end{align}
Here we define 
\begin{equation}
\hat Z_i(\beta_i)=\sum_{n,m} B_{n,m}(\beta_i) (A_i^{\dagger})^n (A_i)^m,i=L,R.
\end{equation}
Because of the special structure of the double-trace correlation function, only $B_{0,n}=B_{n,0}=B_n$ are needed. We refer to the coefficients $B_{n,m}$ as buds. And it must be emphasized that the coefficients $B_n$ cannot be read directly from transfer-matrix eigenstates: as Eq.~\eqref{eq:K2} shows, the $q_{IJ}$ generally cannot be averaged independently in a multitrace calculation.

We also define the $n$th kernel of the double-trace correlation function by
\begin{equation}
\mathcal{K}_n=\la\tr((A_L)^n (A_R^\dagger)^n) \ra.
\end{equation}
A central goal of this work is to evaluate the $n$th bud $B_n$ and kernel $\mathcal K_n$ as $n\to\infty$.

\subsection{The buds}
The preceding leading-order calculation gives
\begin{align}
B_0(\beta)&=Z(\beta), \\
B_1(\beta)&=0, \\
B_2 (\beta) &=\frac{\beta}{2} \frac{\partial}{\partial \beta} Z(\beta).
\end{align}   
Here we set $B_1(\beta)=0$ because  $\mathcal K_1=0$.

For general $B_n$, we use the moment expansion
\begin{equation}
	B_n(\beta)=\sum_{l\geq n} \frac{(-\beta)^l}{l!}\times B'_{n,l},
\end{equation}
where $(-\beta)^l/l!$ comes from expanding $\hat Z=\tr(e^{-\beta \hat H})$ and $B'_{n,l}$ is the contribution to $B_n(\beta)$ from $\tr(\hat H^l)$. 
We can then express $B'_{n,l}$ as
\begin{equation}
B'_{n,l}=\sum_{l_1,\dots,l_n} B''_{n;l_1,\dots,l_n},\sum_i l_i=l-n.
\end{equation}
The coefficient $B''_{n;l_1,\dots,l_n}$ is represented by a chord diagram with open chords:
\begin{align}
B''_{n;l_1,\dots,l_n}&=
\begin{tikzpicture}
\draw [thick,black] (-3,0) -- (-2.5,0);
\draw [thick,black] (-2.5,0) -- (-2.5,1);
\draw [thick,black] (-2.5,0) -- (-2.,0);
\draw (-1.75,0) circle [radius=0.25];
\draw (-1.75,0) node {$l_1$};
\draw [thick,black] (-1.5,0) -- (-0.5,0);
\draw [thick,black] (-1,0) -- (-1,1);
\draw [thick,black] (-1,0) -- (-0.5,0);
\draw  (-0.25,0) circle [radius=0.25];
\draw  (-0.25,0) node {$l_2$};
\draw (0.5,0) node {$\dots$};
\draw [thick,black] (1,0) -- (1.5,0);
\draw [thick,black] (1.5,0) -- (1.5,1);
\draw [thick,black] (1.5,0) -- (2,0);
\draw  (2.25,0) circle [radius=0.25];
\draw  (2.25,0) node {$l_n$};
\draw [thick,black] (2.5,0) -- (3,0);
\end{tikzpicture} 
\end{align}   

Equation~\eqref{eq:LRchord-1} implies that an intersection between an open and a closed chord produces the same factor $q_{IJ}$ as an intersection between two closed chords. However, $\tr(H^l)\neq0$ requires every fermion to appear an even number of times and therefore imposes the additional constraint
\begin{equation}
\label{eq:bud-constrain}
q_{I_1,J} \dots q_{I_n,J}=1,
\end{equation}  
where $I_i$, $i=1,\ldots,n$, label the open chords and $J$ labels any closed chord that intersects all of them. Consequently, Eq.~\eqref{eq:bud-constrain} prevents the $q_{I_i,J}$ from being averaged independently to $q$.

For instance, when $n=3$,
\begin{equation}
\begin{tikzpicture}
\draw (-2,0)--(2,0);
\draw (-1,0) -- (-1,1.5);
\draw (0,0) -- (0,1.5);
\draw (1,0) -- (1,1.5);
\draw (-1.5,0) -- (-1.5,1) -- (1.5,1) -- (1.5,0);
\end{tikzpicture}
= 1	
\end{equation}
and 
\begin{equation}
\begin{tikzpicture}
		\draw (-2,0)--(2,0);
		\draw (-1,0) -- (-1,1.5);
		\draw (0,0) -- (0,1.5);
		\draw (1,0) -- (1,1.5);
		\draw (-1.5,0) -- (-1.5,1) -- (0.5,1) -- (0.5,0);
\end{tikzpicture}
= q=
\begin{tikzpicture}
	\draw (-2,0)--(2,0);
	\draw (-1,0) -- (-1,1.5);
	\draw (0,0) -- (0,1.5);
	\draw (1,0) -- (1,1.5);
	\draw (-1.5,0) -- (-1.5,1) -- (-0.5,1) -- (-0.5,0);
\end{tikzpicture}
.
\end{equation}
These two examples differ qualitatively from conventional single-trace diagrams containing only closed chords. For a given closed chord, a configuration with $k$ intersections with open chords is equivalent to one with $n-k$ intersections. The cyclic symmetry of the trace selects the representation with the smaller number of intersections.

We are interested in the large-$n$ regime that controls the late-time spectral form factor. As $n\to\infty$, the periodic boundary condition of the chord diagram can be replaced effectively by an infinite-boundary condition, which suggests that a naive approximation
\begin{align}
	B'_{n,l}=\sum_{l_1,\dots,l_n} B''_{n;l_1,\dots,l_n}& \simeq  \frac{l}{n} \sum_{l_1,\dots,l_n}
	\begin{tikzpicture}
		\draw [thick,black] (-3,0) -- (-2.5,0);
		\draw [thick,red] (-2.5,0) -- (-2.5,3);
		\draw [thick,black] (-2.5,0) -- (-2.,0);
		\draw (-1.75,0) circle [radius=0.25];
		\draw (-1.75,0) node {$l_1$};
		\draw [thick,black] (-1.5,0) -- (-0.5,0);
		\draw [thick,black] (-1,0) -- (-1,2) -- (3,2);
		\draw [thick,black] (-1,0) -- (-0.5,0);
		\draw  (-0.25,0) circle [radius=0.25];
		\draw  (-0.25,0) node {$l_2$};
		\draw (0.5,0) node {$\dots$};
		\draw [thick,black] (1,0) -- (1.5,0);
		\draw [thick,black] (1.5,0) -- (1.5,1) -- (3,1);
		\draw [thick,black] (1.5,0) -- (2,0);
		\draw  (2.25,0) circle [radius=0.25];
		\draw  (2.25,0) node {$l_n$};
		\draw [thick,black] (2.5,0) -- (3,0);
	\end{tikzpicture}
	\nn \\
	&= \frac{l}{n} \la n-1| T^{l-1} |0\ra.  
\end{align} 

The first open chord is fixed at the first position to remove the cyclic symmetry of the trace, producing the factor $l/n$.  For consistence checking, we have 
\begin{equation}
	\la n| T^n |0\ra =1,
\end{equation}
if $l-n=0$.

In addition, some chords near the two ends of the diagram should not be suppressed by $q^n$. For example,
\begin{equation}
\begin{tikzpicture}
	\draw [thick,black] (-3,0) -- (-2.5,0);
	\draw [thick,red] (-2.5,0) -- (-2.5,3);
	\draw [thick,black] (-2.5,0) -- (-2.,0);
	\draw (-1.75,0) circle [radius=0.25];
	\draw (-1.75,0) node {$l_1$};
	\draw [thick,black] (-1.5,0) -- (-0.5,0);
	\draw [thick,black] (-1,0) -- (-1,2) -- (3,2);
	\draw [thick,black] (-1,0) -- (-0.5,0);
	\draw  (-0.25,0) circle [radius=0.25];
	\draw  (-0.25,0) node {$l_2$};
	\draw (0.5,0) node {$\dots$};
	\draw [thick,black] (1,0) -- (1.5,0);
	\draw [thick,black] (1.5,0) -- (1.5,1) -- (3,1);
	\draw [thick,black] (1.5,0) -- (2,0);
	\draw  (2.25,0) circle [radius=0.25];
	\draw  (2.25,0) node {$l_n$};
	\draw [thick,black] (2.5,0) -- (3,0);
	\draw [thick,green] (-3,2.5) -- (-1.75,2.5) -- (-1.75,0.25);
	\draw [thick,green] (3,0.5) -- (2.25,0.5) -- (2.25,0.25); 
\end{tikzpicture}
=
\begin{tikzpicture}
	\draw [thick,black] (-3,0) -- (-2.5,0);
	\draw [thick,red] (-2.5,0) -- (-2.5,3);
	\draw [thick,black] (-2.5,0) -- (-2.,0);
	\draw (-1.75,0) circle [radius=0.25];
	\draw (-1.75,0) node {$l_1$};
	\draw [thick,black] (-1.5,0) -- (-0.5,0);
	\draw [thick,black] (-1,0) -- (-1,2) -- (3,2);
	\draw [thick,black] (-1,0) -- (-0.5,0);
	\draw  (-0.25,0) circle [radius=0.25];
	\draw  (-0.25,0) node {$l_2$};
	\draw (0.5,0) node {$\dots$};
	\draw [thick,black] (1,0) -- (1.5,0);
	\draw [thick,black] (1.5,0) -- (1.5,1) -- (3,1);
	\draw [thick,black] (1.5,0) -- (2,0);
	\draw  (2.25,0) circle [radius=0.25];
	\draw  (2.25,0) node {$l_n$};
	\draw [thick,black] (2.5,0) -- (3,0);
	\draw [thick,green] (-1.75,0.25) -- (-1.75,2.5) -- (2.25,2.5) -- (2.25,0.25);
\end{tikzpicture}	
\end{equation}
The right-hand side is suppressed by $q^{n-1}$ in the naive approximation. According to Eq.~\eqref{eq:bud-constrain}, however, it should instead be of order $q$ and equal the left-hand side. Our strategy is to naively overcount both these two diagrams and large-$n$ limit picks up the correct answer $\lim_{n\rightarrow \infty} q^k+q^{n-k}\rightarrow q^k$ if $0\leq k \ll n$. Meanwhile, if $k\simeq n$, both $q^k$ and $q^{n-k}$ are subleading in the large $n$ limit. 

The naive approximation must therefore be corrected to
\begin{align}
	B'_{n+1,l}(q)&\simeq \frac{l}{n+1} \sum_{k=0}^{\infty} q^k \oint_{|q_M|=\epsilon}\frac{dq_M}{2\pi i} \frac{1}{q_M^{n+2k+1}} \la n+ k| M T^{l-1} M |k\ra,  n \rightarrow \infty. 
\end{align}
We introduce auxiliary matter insertions to remove redundant chords, as illustrated by
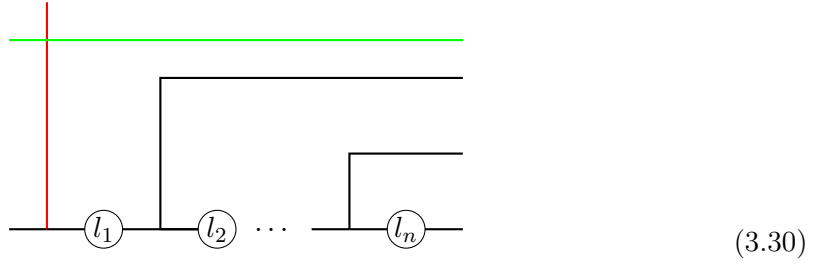
\begin{equation}
\begin{tikzpicture}
	\draw [thick,black] (-3,0) -- (-2.5,0);
	\draw [thick,red] (-2.5,0) -- (-2.5,3);
	\draw [thick,black] (-2.5,0) -- (-2.,0);
	\draw (-1.75,0) circle [radius=0.25];
	\draw (-1.75,0) node {$l_1$};
	\draw [thick,black] (-1.5,0) -- (-0.5,0);
	\draw [thick,black] (-1,0) -- (-1,2) -- (3,2);
	\draw [thick,black] (-1,0) -- (-0.5,0);
	\draw  (-0.25,0) circle [radius=0.25];
	\draw  (-0.25,0) node {$l_2$};
	\draw (0.5,0) node {$\dots$};
	\draw [thick,black] (1,0) -- (1.5,0);
	\draw [thick,black] (1.5,0) -- (1.5,1) -- (3,1);
	\draw [thick,black] (1.5,0) -- (2,0);
	\draw  (2.25,0) circle [radius=0.25];
	\draw  (2.25,0) node {$l_n$};
	\draw [thick,black] (2.5,0) -- (3,0);
	\draw [thick,green] (-3,2.5) -- (3,2.5);
\end{tikzpicture}		
\end{equation}
The parameter $q_M$ is the factor associated with intersections between the auxiliary matter chord and the $H$-chords.

The resulting effective large-$n$ bud is
\begin{align}
	B_{n+1}(\beta;q)&\simeq \frac{-\beta}{n+1} \sum_{k=0}^{\infty} q^k \oint_{|q_M|=\epsilon}\frac{dq_M}{2\pi i} \frac{1}{q_M^{n+2k+1}} \la n+ k| M e^{-\beta T} M |k\ra,  n \rightarrow \infty. 
\end{align}
We notice that for any function $f(T)$, propagation through the complete matter region is
\begin{equation}\label{eq:full-matter-region}
\la m|M f(T)M|\ell\ra
=\sum_{i=0}^{\min(m,\ell)}
\genfrac{[}{]}{0pt}{}{\ell}{i}_{q}
(q_M^2;q)_i q_M^{m+\ell-2i}
\la m-i|f(T)|\ell-i\ra ,
\end{equation}
where
\begin{equation}
\genfrac{[}{]}{0pt}{}{\ell}{i}_{q}
=\frac{(q;q)_\ell}{(q;q)_i(q;q)_{\ell-i}}.
\end{equation}
This formula treats the two matter endpoints as one bilocal insertion and,
in particular, retains the $H$-chords that pass above the entire matter
chord.

We now simplify the bud in four steps.
\begin{enumerate}
\item Set $m=n+k$, $\ell=k$, and $f(T)=e^{-\beta T}$ in
Eq.~\eqref{eq:full-matter-region}.  The $q_M$ contour selects the coefficient
of $q_M^{n+2k}$.  Since
\begin{equation}
\left[q_M^{2i}\right](q_M^2;q)_i
=(-1)^i q^{\binom{i}{2}},
\end{equation}
the quantity multiplying $-\beta/(n+1)$ in the bud becomes
\begin{align}
\mathcal C_n(\beta;q)
=\sum_{k=0}^{\infty}q^k\sum_{i=0}^{k}
(-1)^i q^{\binom{i}{2}}
\genfrac{[}{]}{0pt}{}{k}{i}_{q}
\la n+k-i|e^{-\beta T}|k-i\ra .
\label{eq:Cn-after-contour}
\end{align}

\item Write a term in the exponential as $T^{n+2s}$.  A path from
$|r\ra$ to $|n+r\ra$ then contains $n+s$ chord openings and $s$ chord
closings.  In the large-$n$ tail,
\begin{equation}
[l]_q=\frac{1-q^l}{1-q}\longrightarrow\frac{1}{1-q}.
\end{equation}
For fixed $s$ (more generally, $s=o(n)$), endpoint-sensitive paths are
subleading.  The leading bulk matrix element is therefore
\begin{equation}\label{eq:asymptotic-path-count}
\la n+r|T^{n+2s}|r\ra
\simeq \binom{n+2s}{s}\frac{1}{(1-q)^s},
\end{equation}
independently of $r$.

\item Because Eq.~\eqref{eq:asymptotic-path-count} is independent of
$r=k-i$ at leading order, the inner sum in
Eq.~\eqref{eq:Cn-after-contour} is evaluated by the finite $q$-binomial
theorem:
\begin{equation}
\sum_{i=0}^{k}(-1)^i q^{\binom{i}{2}}
\genfrac{[}{]}{0pt}{}{k}{i}_{q}
=(1;q)_k=\delta_{k0}.
\end{equation}
Thus all redundant $k>0$ sectors cancel at leading order, and the thermal
series resums as
\begin{align}
\mathcal C_n(\beta;q)
&\simeq
\sum_{s=0}^{\infty}
\frac{(-\beta)^{n+2s}}{(n+2s)!}
\binom{n+2s}{s}\frac{1}{(1-q)^s}
\nn \\
&=(-1)^n(1-q)^{n/2}
I_n\left(\frac{2\beta}{\sqrt{1-q}}\right),
\label{eq:Cn-bessel}
\end{align}
where $I_n$ is the modified Bessel function of the first kind, defined in
Appendix~\ref{app:modified-bessel}.

\item Substitution into the bud gives the useful intermediate form
\begin{equation}
B_{n+1}(\beta;q)
\simeq
(-1)^{n+1}\frac{\beta}{n+1}(1-q)^{n/2}
I_n\left(\frac{2\beta}{\sqrt{1-q}}\right).
\end{equation}
Using $I_n(x)-I_{n+2}(x)=2(n+1)I_{n+1}(x)/x$ and
$I_{n+2}(x)/I_n(x)=O(n^{-2})$ at fixed $x$, we finally obtain
\begin{equation}\label{eq:simplified-large-n-bud}
\boxed{
B_{n+1}(\beta;q)
\simeq
(-1)^{n+1}(1-q)^{\frac{n+1}{2}}
I_{n+1}\left(\frac{2\beta}{\sqrt{1-q}}\right)}
,\qquad n\to\infty .
\end{equation}
\end{enumerate}

As checks, Eq.~\eqref{eq:simplified-large-n-bud} reduces at $q=0$ to
$B_{n+1}=(-1)^{n+1}I_{n+1}(2\beta)$, and its first small-$\beta$ term is
$(-\beta)^{n+1}/(n+1)!$, as required by the diagram containing only the
$n+1$ open chords. And
\begin{equation}\label{eq:minimal-bud-coefficient}
B^{\prime}_{n,n}(q)=1.
\end{equation}
This equality is exact: at $l=n$ the diagram contains only the $n$ open
chords, so there are no closed chords or intersection weights and hence no
$q$ dependence.

\subsection{The kernels}
\label{sec:kernels}
We begin with the simplest kernel:
\begin{align}
\mathcal K_3=&\la \tr(A_L A_L A_L A_R^\dagger A_R^\dagger A_R^\dagger ) \ra \nn \\
=&\mathcal J^6 \sum_{I_i}  \la \tr(A_L^{I_1} A_L^{I_2} A_L^{I_3} (A_R^\dagger)^{I_4} (A_R^\dagger)^{I_5} (A_R^\dagger)^{I_6} ) \ra
\nn \\
=&\mathcal J^6 \sum_{I_i} \tr(I_{LR}^{I_1}I_{LR}^{I_2}I_{LR}^{I_3}) \nn \\ &(1+q_{I_1,I_2}+q_{I_2,I_3}+q_{I_1,I_2}q_{I_1,I_3}+q_{I_1,I_2}q_{I_2,I_3}+q_{I_1,I_2}q_{I_1,I_3}q_{I_2,I_3})
\end{align}
For the expression
\begin{equation}
\tr(I_{LR}^{I}I_{LR}^{J}I_{LR}^{K})= \tr( \Psi_I^L\Psi_J^L \Psi_K^L) \tr( \Psi_I^R\Psi_J^R \Psi_K^R),   \
\end{equation}
to be nonzero, every fermion $\psi_i$ in $\la\Psi_I\Psi_J\Psi_K\ra_\psi$ must be paired. Since the left and right factors are identical, it is sufficient to analyze one side.

Defining $m_{IJ}=|I\cap J|$ gives
\begin{align}
	m_{IJ}+m_{IK}&=p, \\
	m_{IJ}+m_{JK}&=p, \\
	m_{IK}+m_{JK}&=p.
\end{align}
The unique solution is $m_{IJ}=m_{IK}=m_{JK}=p/2$, and every fermion $\psi_i$ appears exactly twice. If $p/2$ is even, then
\begin{equation}
q_{I_1,I_2}=q_{I_1,I_3}=q_{I_2,I_3}=1
\end{equation}
and 
\begin{equation}
\mathcal{K}_3=6	\mathcal J^6 \binom{N}{p/2} \binom{N-p/2}{p/2} \binom{N-p}{p/2}.
\end{equation}
Otherwise,
\begin{equation}
	q_{I_1,I_2}=q_{I_1,I_3}=q_{I_2,I_3}=-1,
\end{equation}
which gives $\mathcal K_3=0$. We have thus reproduced the $n=3$ result of Ref.~\cite{Berkooz:2020fvm}.

The general kernels $\mathcal K_{n\geq4}$ are more involved. We write
\begin{align}
	\label{eq:kerneln}
	\mathcal{K}_n&=\mathcal J^{2n} \sum_{I_i} \tr(I_{LR}^{I_1}I_{LR}^{I_2}\dots I_{LR}^{I_n})  F_n(q)
\end{align}
Here we take the double-scaled and large-$n$ limits. The factor $F_n(q)$ represents the contribution from intersections between cross-trace chords, while
\begin{equation}
\tr(I_{LR}^{I_1}I_{LR}^{I_2}\dots I_{LR}^{I_n})=\sum_{\{I_i\}}\tr(\Psi^L_{I_1} \dots \Psi^L_{I_n} )\tr(\Psi^R_{I_1} \dots \Psi^R_{I_n} )	
\end{equation}
denotes the cross-trace ladder diagram. Requiring $\tr(I_{LR}^{I_1}I_{LR}^{I_2}\cdots I_{LR}^{I_n})$ to be nonzero imposes the additional constraint
\begin{equation}
	\label{eq:kernel-constrain}
	\prod_{i\neq j}q_{I_i,I_j}=1.
\end{equation}
One might expect this condition to become less important as $n\to\infty$. However, Eq.~\eqref{eq:kernel-constrain} induces a cyclic symmetry that persists in this limit and produces an overall factor of $n$ in Eq.~\eqref{eq:kerneln}.

The remaining factor $F_n(q)$ still cannot be evaluated as an ordinary chord-diagram sum. A fermion in $\tr(\Psi^X_{I_1}\cdots\Psi^X_{I_n})$, with $X=L,R$, may appear $2k$ times for any $k\geq1$. Nevertheless, when $np\ll N$, configurations in which every $\psi_i$ appears exactly twice dominate at large $N$. The overlaps $m_{ij}=|I_i\cap I_j|$ are then constrained by
\begin{equation}
\label{eq:m-equations}
\begin{aligned}
	m_{12}+m_{13}+...m_{1n}&=p \\
	m_{21}+m_{23}+...m_{2n}&=p \\
	... & \\
	m_{n2}+m_{n3}+...m_{n,n-1}&=p.
\end{aligned}
\end{equation}
There are $n(n-1)/2$ variables but only $n$ equations, leaving $n(n-3)/2$ independent integers. For a given solution, define
\begin{equation}
	m_1=m_{12},m_2=m_{13},\dots m_x=m_{i<j} \dots m_{\frac{n(n-1)}{2}}=m_{n-1,n}.
\end{equation} 
Then
\begin{align}
	\mathcal{K}_n \sim  \sum_{\text{solutions}} \prod_i \binom{N-\sum_{j<i} m_j}{m_i}.
\end{align}
If $\frac{p}{n-1}>1$, the saddle point is roughly
\begin{equation}
	m=\frac{p}{n-1}.
\end{equation}
For $n=2,3$, this saddle coincides with the unique solution of Eqs.~\eqref{eq:m-equations}. 

We are interested instead in the region $np>N$, where Eqs.~\eqref{eq:m-equations} no longer constrain the $m_{ij}$, although Eq.~\eqref{eq:kernel-constrain} continues to affect the kernel. 
We therefore treat the cross-trace chords as ordinary chords with intersection weight 
\begin{equation}
	\la q_{IJ} \ra=q
\end{equation}
and additional corrections from
the global constraint \eqref{eq:kernel-constrain}.

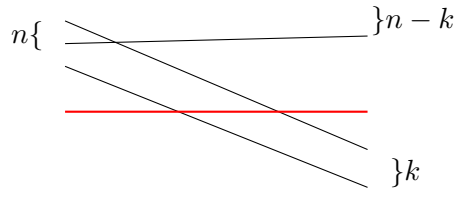
\begin{figure}
\label{fig:kernel}
\centering
\begin{tikzpicture}
\draw (-2,1.5) -- (2,1.5);
\draw (-2,1.2) -- (2,-0.5);
\draw (-2,0.9) -- (2,1);
\draw  (-2,0.6) -- (2, -1);	
\draw [red, thick](-2,0) -- (2,0);
\draw  (-2.5,1) node {$n\large \{ $};
\draw  (2.6,1.2) node {$\large \} n-k $};
\draw  (2.5,-0.8) node {$\large \} k $};
\end{tikzpicture}
\caption{The reference chord is labeled by red color.}
\end{figure}

Let's derive the naive version of $F_{n+1}(\{q_{I_i,I_j}\})$.
First, fix one cross-trace chord as a reference, removing the cyclic symmetry and producing a factor of $n+1$. Divide the remaining $n$ chords into those that intersect the reference chord and those that do not, as shown as Fig. \ref{fig:kernel}. If $k$ chords intersect it, this gives a factor $q^k\binom{n}{k}_q$. And self-intersections within the two groups contribute $\frac{(q;q)_k}{(1-q)^k}\frac{(q;q)_{n-k}}{(1-q)^{n-k}}$. 

Thus,
\begin{align}
F_{n+1}(q)
&=(n+1) \sum_{k=0}^n  \la A^{k} M A^{n-k} (A^\dagger)^n M \ra, q_m\rightarrow q.
\nn \\
&=(n+1)\sum_{k}^{\infty} q^k \frac{(q;q)_{n}}{(q;q)_{k} (q;q)_{n-k}} \frac{(q;q)_k}{(1-q)^k} \frac{(q;q)_{n-k}}{(1-q)^{n-k}}
\nn \\
&\simeq (n+1)\sum_{k=0}^\infty q^k \frac{(q;q)_n}{(1-q)^n}\nn \\
&=\frac{(n+1)}{1-q} \frac{(q;q)_n}{(1-q)^n}.  	
\end{align}

Again, there are $q^n$ suppressed terms which are enlarged by Eq. \eqref{eq:kernel-constrain}. We continue our overcounting trick to calculate the buds to find the corrections from Eq. \eqref{eq:kernel-constrain}.  And $F_{n+1}(q)$ are corrected to be 
\begin{align}
F_{n+1}(q)
&=(n+1) \sum_{k=0}^n \sum_{l=0}^n \oint_{|q_m|=\epsilon} \frac{dq_m}{2\pi i} q^{k+l} \frac{1}{q_m^{2l+k+1}}  \la l| A^{k} M A^{n-k} (A^\dagger)^n M |l\ra	.
\end{align}

We have 
\begin{equation}
	\langle l|A^kMA^{n-k}(A^\dagger)^nM|l\rangle
	=q_m^{2l+k}\frac{[l+n]_q !}{[l]_q!},
\end{equation} 
therefore
\begin{align}
	F_{n+1}(q)
	&=(n+1)\sum_{k=0}^{n}\sum_{l=0}^{n}
	q^{l+k}\frac{[l+n]_q !}{[l]_q !}
	\notag\\
	&=(n+1)[n+1]_q\sum_{l=0}^{n}
	q^l \frac{[l+n]_q !}{[l]_q !}.
	\label{eq:weighted-sums}
\end{align}
Here \(\sum_{k=0}^{n}q^k=[n+1]_q\).  The remaining ratio is
\begin{equation}
	\frac{\qfac{l+n}}{\qfac{l}}
	=\qfac{n}\qbinom{n+l}{n}.
\end{equation}
The finite \(q\)-hockey-stick identity reads
\begin{equation}
	\sum_{l=0}^{m}q^l\qbinom{r+l}{r}
	=\qbinom{r+m+1}{r+1}.
\end{equation}
Setting \(r=m=n\) therefore gives
\begin{align}
	\sum_{l=0}^{n}q^l\frac{\qfac{l+n}}{\qfac{l}}
	&=\qfac{n}\sum_{l=0}^{n}q^l\qbinom{n+l}{n}
	\notag\\
	&=\qfac{n}\qbinom{2n+1}{n+1}.
\end{align}
Combining this with the \(k\)-sum yields
\begin{align}
	F_{n+1}(q)
	&=(n+1)[n+1]_q\qfac{n}\qbinom{2n+1}{n+1}
	\notag\\
	&=(n+1)\qfac{n+1}\qbinom{2n+1}{n+1}
	\notag\\
	&=(n+1)\frac{\qfac{2n+1}}{\qfac{n}}\nn \\
	&=\frac{(n+1)(q;q)_{2n+1}}
	{(1-q)^{n+1}(q;q)_n}.
	\label{eq:main}
\end{align}
In the large-$n$ limit, $(q;q)_{2n+1}=(q;q)_{n}=(q;q)_{\infty}$, such that
\begin{equation}
\lim_{n\rightarrow \infty}F_{n+1}(q)= \frac{n+1}{(1-q)^{n+1}}.
\end{equation}

\begin{figure}[htbp]
	\centering
	\begin{tikzpicture}[line cap=round,line join=round]
		\def\cylradius{1.45}
		\def\cylheight{2.4}
		\def\ellipseheight{0.4}
		
		\foreach \xcenter in {-2.15,2.15}{
			\begin{scope}[shift={(\xcenter,0)}]
				\fill[blue!7] (-\cylradius,0)
				rectangle (\cylradius,\cylheight);
				\fill[blue!12] (0,\cylheight)
				ellipse [x radius=\cylradius,y radius=\ellipseheight];
				\fill[blue!5] (0,0)
				ellipse [x radius=\cylradius,y radius=\ellipseheight];
				
				\draw[dashed,gray!70,thick]
				(-\cylradius,0)
				arc[start angle=180,end angle=0,
				x radius=\cylradius,y radius=\ellipseheight];
				\draw[thick] (-\cylradius,0) -- (-\cylradius,\cylheight);
				\draw[thick] ( \cylradius,0) -- ( \cylradius,\cylheight);
				\draw[thick] (0,\cylheight)
				ellipse [x radius=\cylradius,y radius=\ellipseheight];
				\draw[thick]
				(-\cylradius,0)
				arc[start angle=180,end angle=360,
				x radius=\cylradius,y radius=\ellipseheight];
			\end{scope}
		}
		
		\draw[->,very thick,red!70!black]
		(-2.15,\cylheight) --
		(-2.15,\cylheight+0.3);
		\draw[->,very thick,red!70!black]
		(-2.15,0) --
		(-2.15,0.3);
		\node[font=\small] at (-2.15,\cylheight+0.75)
		{top: \(\rightarrow\)};
		\node[font=\small] at (-2.15,-0.75)
		{bottom: \(\rightarrow\)};
		
		\draw[->,very thick,red!70!black]
		(2.15,\cylheight) --
		(2.15,\cylheight-0.3);
		\draw[->,very thick,red!70!black]
		(2.15,0) --
		(2.15,+0.3);
		\node[font=\small] at (2.15,\cylheight+0.75)
		{top: \(\leftarrow\)};
		\node[font=\small] at (2.15,-0.75)
		{bottom: \(\rightarrow\)};
	\end{tikzpicture}
	\caption{The two relative orientations for gluing the top and bottom disks
		to form a cylinder.  The disk orientations agree on the left
		\((\rightarrow,\rightarrow)\) and are opposite on the right
		\((\rightarrow,\leftarrow)\), where each ordered pair lists the bottom
		orientation first.}
	\label{fig:cylinder-orientations}
\end{figure}
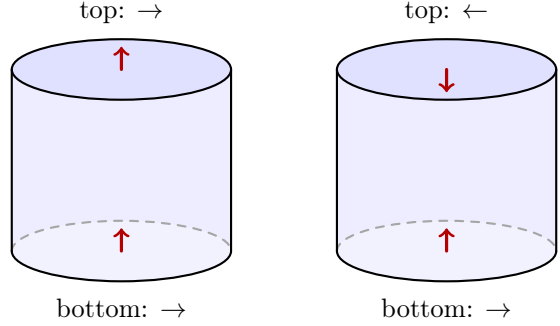

However, this is not the final result either. We notice that there is no prior orientation to glue to two disks together and form a cylinder, as shown as Fig. \ref{fig:cylinder-orientations}. This orientation can also be identified with the time directions in these two boundary theories. Such that the physical kernel should sum the two possibilities. And our above calculation corresponds to the orientation $(\rightarrow,\rightarrow)$. For the another orientation $(\rightarrow,\leftarrow)$, it is not only the permutation of the order in one trace but also changes any $\Psi_I=\psi_{i_1}...\psi_{i_p}$ to   
\begin{equation}
\Psi'_I=\psi_{i_p}...\psi_{i_i}=(-1)^{p(p-1)/2}\Psi_I=(-1)^{p/2}\Psi_I.
\end{equation}
And the overall sign for $(\rightarrow,\leftarrow)$ is $(-1)^{(n+1)p/2}$ for $F_{n+1}(q)$. Finally, we obtain that
\begin{equation}
\lim_{n\rightarrow \infty}F_{n+1}(q)=
\begin{cases}
\frac{2(n+1)}{(1-q)^{n+1}}, \frac{(n+1)p}{2} \in \text{Even}, \\
0, \frac{(n+1)p}{2} \in \text{Odd}.	
\end{cases}
\end{equation}

On the other hand, Ref.~\cite{Cotler:2016fpe} showed that
\begin{align}
\tr(I_{LR}^n)&=\mathcal J^{2n}\sum_{I_1,...,I_n} \tr(I_{LR}^{I_1}I_{LR}^{I_2}\dots I_{LR}^{I_n}) 
\nn \\
&= \mathcal J^{2n} 2^{-N} \sum_{x_i=\pm 1} (\sum_{i_1<i_2<\dots <i_p} x_{i_1}\dots x_{i_p})^n \nn \\
&=\mathcal J^{2n} 2^{-N} \sum_{m=0}^N \binom{N}{m} (\sum_{k=0}^p \binom{m}{k} \binom{N-m}{p-k}(-1)^k)^n.
\end{align}
Substituting $\mathcal J^2=\binom{N}{p}^{-1}$ gives
\begin{align}
\tr(I_{LR}^n)&=2^{-N} \sum_{m=0}^N \binom{N}{m} \alpha(N,m,p)^n 
 \\ 
\alpha(N,m,p)
&=\frac{(N-q)!(N-m)!}{N!}\frac{_2 F_1(-m,-q,N-m-q+1,-1)}{\Gamma(N-m-q+1)} 
\end{align}
Ref.~\cite{Cotler:2016fpe} argues that, for sufficiently large $n$, $\tr(I_{LR}^n)$ is dominated by the saddles with $\alpha(N,m,p)=1$ at $m=0,N$. Hence
\begin{equation}
\tr(I_{LR}^n)\simeq2\times2^{-N} \equiv c_N.	
\end{equation}

Combining these results gives
\begin{equation}
\label{eq:simplified-large-n-kernel}
	\lim_{n\rightarrow \infty}\mathcal K_{n+1}(q)=
	\begin{cases}
		\frac{2 c_N (n+1)}{(1-q)^{n+1}}, \frac{(n+1)p}{2} \in \text{Even} \\
		0, \frac{(n+1)p}{2} \in \text{Odd}	
	\end{cases}.
\end{equation}

\subsection{Ramp}
The higher-order Cactus correction to the spectral form factor is
\begin{equation}
\Delta Z(\beta_L,\beta_R)=\sum_{n\geq \Lambda} B_{n+1}(\beta_L) \mathcal{K}_{n+1} B_{n+1}(\beta_R)
\end{equation}
where $\Lambda$ is a cutoff, and
\begin{equation}
	\beta_L=\beta+i t, \beta_R=\beta -i t .
\end{equation}
We are interested in the singular terms that are insensitive to $\Lambda$. We therefore write
\begin{equation}
	Z_{s}(\beta_L,\beta_R)\sim \sum_{n=0}^\infty B_{n+1}(\beta_L) \mathcal{K}_{n+1} B_{n+1}(\beta_R)
\end{equation}
where the subscript $s$ distinguishes it from the original $Z(\beta_L,\beta_R)$. We use $\sim$ rather than $=$ because the two sides agree only in their singular terms.

For brevity, define
\begin{equation}
c=1-q,\qquad x_L=\frac{2\beta_L}{\sqrt c},
\qquad x_R=\frac{2\beta_R}{\sqrt c}.
\end{equation}
Assuming that $4|p$,
substituting Eqs.~\eqref{eq:simplified-large-n-bud} and
\eqref{eq:simplified-large-n-kernel} into the Cactus sum cancels all explicit
powers of $c$:
\begin{equation}
Z_s(\beta_L,\beta_R)
\sim 2c_N\sum_{n=0}^{\infty}
(n+1) I_{n+1}(x_L)I_{n+1}(x_R).
\end{equation}

The remaining series can be summed exactly.  Applying
$2mI_m(x)/x=I_{m-1}(x)-I_{m+1}(x)$ for both arguments makes the partial sum
telescope and gives
\begin{equation}\label{eq:weighted-bessel-sum}
\sum_{m=1}^{\infty}m I_m(x)I_m(y)
=\frac{xy}{2(x+y)}
\left[I_0(x)I_1(y)+I_1(x)I_0(y)\right].
\end{equation}
The convergence and the telescoping argument are given in
Appendix~\ref{app:modified-bessel}.
The apparent pole at $x+y=0$ is removable.  Hence
\begin{equation}\label{eq:Zs-singular-closed}
\boxed{
Z_s^{\mathrm{sing}}(\beta_L,\beta_R)
=2c_N\frac{x_Lx_R}{2(x_L+x_R)}
\left[I_0(x_L)I_1(x_R)+I_1(x_L)I_0(x_R)\right]},
\end{equation}
for $4\mid p$.

Now set
\begin{equation}
\beta_L=\beta+it,\qquad\beta_R=\beta-it,\qquad
a=\frac{2\beta}{\sqrt c},\qquad \tau=\frac{2t}{\sqrt c}.
\end{equation}
Thus $x_L=a+i\tau$ and $x_R=a-i\tau$.  The large-$|\tau|$ asymptotics of
Eq.~\eqref{eq:weighted-bessel-sum} is
\begin{equation}
\sum_{m=1}^{\infty}m I_m(a+i\tau)I_m(a-i\tau)
=\frac{|\tau|}{\pi}\frac{\sinh(2a)}{2a}+O(1).
\end{equation}
The nonanalytic late-time term is therefore
\begin{equation}\label{eq:Zs-linear-ramp}
\boxed{
Z_s^{\mathrm{sing}}(\beta+it,\beta-it)
=2c_N
\frac{\sinh\left(4\beta/\sqrt{1-q}\right)}{2\pi\beta}
|t|+O(1)}
\end{equation}
for $4 \mid p$.

For $4\nmid p$, recall that $p$ is even, so necessarily
$p=2\pmod 4$.  The condition in
Eq.~\eqref{eq:simplified-large-n-kernel} then keeps only even values of
$n+1$.  Consequently,
\begin{equation}
Z_s(\beta_L,\beta_R)
\sim 2c_N\sum_{m=1}^{\infty}
2m I_{2m}(x_L)I_{2m}(x_R).
\end{equation}
The even-order sum can be obtained by a parity projection:
\begin{align}
\sum_{m=1}^{\infty}2m I_{2m}(x)I_{2m}(y)
&=\frac{1}{2}\sum_{k=1}^{\infty}
k I_k(x)\left[I_k(y)+I_k(-y)\right]
\notag\\
&=\frac{xy}{2}
\frac{xI_0(x)I_1(y)-yI_1(x)I_0(y)}{x^2-y^2}.
\label{eq:even-weighted-bessel-sum}
\end{align}
Thus the singular part has the closed form
\begin{equation}
\boxed{
Z_s^{\mathrm{sing}}(\beta_L,\beta_R)
=c_Nx_Lx_R
\frac{x_LI_0(x_L)I_1(x_R)-x_RI_1(x_L)I_0(x_R)}
{x_L^2-x_R^2}},
\qquad 4\nmid p .
\label{eq:Zs-even-singular-closed}
\end{equation}
As in Eq.~\eqref{eq:Zs-singular-closed}, the apparent poles are
removable.

For $x=a+i\tau$ and $y=a-i\tau$, the first term in the parity
projection is the sum in Eq.~\eqref{eq:weighted-bessel-sum}, while the
alternating term is bounded:
\begin{equation}
\sum_{k=1}^{\infty}(-1)^k k
I_k(a+i\tau)I_k(a-i\tau)
=\frac{\cos(2\tau)}{2\pi}+O(|\tau|^{-1}).
\end{equation}
It therefore does not generate another nonanalytic term.  The
even-order sum has the late-time behavior
\begin{equation}
\sum_{m=1}^{\infty}2m
I_{2m}(a+i\tau)I_{2m}(a-i\tau)
=\frac{|\tau|}{2\pi}\frac{\sinh(2a)}{2a}+O(1),
\end{equation}
and hence
\begin{equation}\label{eq:Zs-linear-ramp-p-2-mod-4}
\boxed{
Z_s^{\mathrm{sing}}(\beta+it,\beta-it)
=c_N
\frac{\sinh\left(4\beta/\sqrt{1-q}\right)}{2\pi\beta}
|t|+O(1)}
\end{equation}
for $4\nmid p$.  The nonanalytic ramp is therefore one half of the
$4\mid p$ result in Eq.~\eqref{eq:Zs-linear-ramp}.

The difference between $4\mid p$ and $4\nmid p$ reproduces the symmetry factor found in Ref.~\cite{Cotler:2016fpe}.
Moreover, the temperature-dependent factor
\begin{equation}
\frac{\sinh\left(4\beta/\sqrt{1-q}\right)}{2\pi\beta}=\frac{1}{2\pi} \int_{0}^{\pi} \frac{2\sin \theta}{\sqrt{1-q}} e^{-\frac{4\beta \cos\theta}{\sqrt{1-q}}}d\theta =\frac{1}{2\pi} \int_{E_{min}}^{E_{max}} e^{-2\beta E} dE	
\end{equation}
agrees with the semiclassical prediction of Ref.~\cite{Goel:2023svz}.

Collecting the two cases, we obtain
\begin{equation}
\boxed{
Z_s^{\mathrm{sing}}(\beta+it,\beta-it)
=s_p c_N
\frac{|t|}{2\pi} \int_{E_{min}}^{E_{max}} e^{-2\beta E} dE+O(1)}
\end{equation}
where
\begin{equation}
	s_p=\begin{cases}
		2,4 \mid p, \\
		1,4 \nmid p.
	\end{cases}
\end{equation}

\section{Conclusion and discussion}

In this work, we derived the late-time ramp of the finite-temperature spectral form factor in double-scaled SYK directly from the large-$n$ tail of Cactus diagrams. The main obstacle had no analogue in the usual single-trace problem: the intersection weights $q_{IJ}$ were correlated by the trace constraints and could not be averaged independently to $q$. Instead of attempting to evaluate every finite-order Cactus diagram, we isolated the nonanalytic part of the resummed series. Our central idea was that the ramp is controlled by the large-$n$ tail and is unchanged by finite modifications of the low-order terms. This reduced the problem to determining the asymptotic contribution with a large number of cross-trace pairings.

We factorized the contribution at order $n$ into two buds and a kernel. For fixed $0\leq q<1$, the large-$n$ bud is
\begin{equation}
B_n(\beta;q)\simeq(-1)^n(1-q)^{n/2}
I_n\left(\frac{2\beta}{\sqrt{1-q}}\right),
\end{equation}
while the kernel behaves as
\begin{equation}
\mathcal K_n(q)\simeq
\begin{cases}
\displaystyle \frac{2c_N n}{(1-q)^n},&np/2\in \text{Even},\\[2mm]
0,&np/2\in \text{Odd}.
\end{cases}
\end{equation}
When $4\mid p$, every $n$ contributes; when $p=2\pmod 4$, only even $n$ contribute. This is the microscopic origin of the symmetry factor in the final answer.

Combining the buds and the kernel cancels the explicit powers of $1-q$ and reduces the Cactus sum to a weighted Bessel series that can be evaluated in closed form. Its nonanalytic late-time part is
\begin{equation}
\boxed{
Z_s^{\mathrm{sing}}(\beta+it,\beta-it)
=s_p c_N
\frac{|t|}{2\pi} \int_{E_{min}}^{E_{max}} e^{-2\beta E} dE+O(1)}
\end{equation}
where
\begin{equation}
s_p=\begin{cases}
	2,4 \mid p, \\
	1,4 \nmid p.
     \end{cases}
\end{equation}
The result reproduced the linear ramp expected from random-matrix theory, the semiclassical temperature dependence and the symmetry factor. It therefore provided a direct chord-diagram derivation of the universal two-level correlations that are invisible in the naive perturbative expansion of the (double scaled) SYK spectral form factor.

The present calculation did not produce the plateau. This is naturally expected because the plateau is intrinsically a finite-$N$ effect. For a finite-dimensional Hilbert space, the spectrum is discrete, and at sufficiently late times the off-diagonal phases in the spectral form factor dephase. The remaining diagonal terms are set by the finite-dimensional trace normalization and by
\begin{equation}
\left\langle\sum_m e^{-2\beta E_m}\right\rangle
=\left\langle Z(2\beta)\right\rangle,
\end{equation}
which determines the plateau height. By contrast, the double-scaled limit sends $N,p\to\infty$ at fixed $\lambda=2p^2/N$ and replaces the microscopic discrete spectrum by a continuous spectral density. The large-$n$ approximation used in this work retained the contribution responsible for the ramp but not the detailed level discreteness required for the plateau.

\acknowledgments
I sincerely thank Hong Liu for suggesting and discussing the topics about double scaled SYK model. AI was used in this work to simplify complicated formulas.

\appendix
\section{Modified Bessel functions and the weighted sums}
\label{app:modified-bessel}

For a nonnegative integer $n$, the modified Bessel function of the first
kind is defined by the everywhere-convergent power series
\begin{equation}
I_n(z)=\sum_{r=0}^{\infty}
\frac{1}{r!(n+r)!}\left(\frac{z}{2}\right)^{n+2r},
\qquad z\in\mathbb C,
\label{eq:appendix-In-definition}
\end{equation}
and $I_{-n}(z)=I_n(z)$.  It is the solution regular at the origin of
\begin{equation}
z^2 I_n''(z)+zI_n'(z)-(z^2+n^2)I_n(z)=0.
\end{equation}
The defining series immediately implies the parity relation
\begin{equation}
I_n(-z)=(-1)^n I_n(z),
\label{eq:appendix-In-parity}
\end{equation}
and term-by-term comparison gives the recurrence
\begin{equation}
I_{n-1}(z)-I_{n+1}(z)=\frac{2n}{z}I_n(z),
\qquad n\geq1,
\label{eq:appendix-In-recurrence}
\end{equation}
where the value at $z=0$ is understood by continuity.

We first prove the weighted sum in
Eq.~\eqref{eq:weighted-bessel-sum}.  For fixed complex $x$ and $y$, define
\begin{equation}
T_m(x,y)=I_m(x)I_{m+1}(y)+I_{m+1}(x)I_m(y).
\end{equation}
Applying Eq.~\eqref{eq:appendix-In-recurrence} once with argument $x$ and
once with argument $y$ gives
\begin{align}
T_{m-1}(x,y)-T_m(x,y)
&=\left[I_{m-1}(x)-I_{m+1}(x)\right]I_m(y)
\notag\\
&\quad+I_m(x)\left[I_{m-1}(y)-I_{m+1}(y)\right]
\notag\\
&=2m\left(\frac{1}{x}+\frac{1}{y}\right)I_m(x)I_m(y).
\label{eq:appendix-telescoping-step}
\end{align}
Consequently, if $xy(x+y)\neq0$, the finite sum telescopes:
\begin{align}
\sum_{m=1}^{M}mI_m(x)I_m(y)
&=\frac{xy}{2(x+y)}
\left[T_0(x,y)-T_M(x,y)\right].
\label{eq:appendix-finite-weighted-sum}
\end{align}
The power series in Eq.~\eqref{eq:appendix-In-definition} gives the bound
\begin{equation}
|I_n(z)|\leq
\frac{(|z|/2)^n}{n!}\exp\left(\frac{|z|^2}{4}\right).
\end{equation}
It follows both that the weighted series converges absolutely and locally
uniformly in $(x,y)$ and that $T_M(x,y)\to0$ as $M\to\infty$.  Taking this
limit in Eq.~\eqref{eq:appendix-finite-weighted-sum}, with
$T_0=I_0(x)I_1(y)+I_1(x)I_0(y)$, proves
\begin{equation}
\sum_{m=1}^{\infty}mI_m(x)I_m(y)
=\frac{xy}{2(x+y)}
\left[I_0(x)I_1(y)+I_1(x)I_0(y)\right].
\end{equation}
Although the derivation temporarily assumed $xy(x+y)\neq0$, local uniform
convergence makes the left-hand side entire in both variables.  The formula
therefore extends by continuity to $x=0$, $y=0$, and $x+y=0$; in particular,
the apparent pole at $x+y=0$ is removable.

The even-order identity in
Eq.~\eqref{eq:even-weighted-bessel-sum} now follows by parity projection.
Using Eq.~\eqref{eq:appendix-In-parity} and absolute convergence,
\begin{align}
\sum_{m=1}^{\infty}2mI_{2m}(x)I_{2m}(y)
&=\frac12\sum_{k=1}^{\infty}kI_k(x)
\left[I_k(y)+I_k(-y)\right]
\notag\\
&=\frac12\left[S(x,y)+S(x,-y)\right],
\label{eq:appendix-even-projection}
\end{align}
where
\begin{equation}
S(x,y)=\sum_{k=1}^{\infty}kI_k(x)I_k(y).
\end{equation}
Substituting the weighted sum just proved gives
\begin{align}
\frac12\left[S(x,y)+S(x,-y)\right]
&=\frac{xy}{4}\left[
\frac{I_0(x)I_1(y)+I_1(x)I_0(y)}{x+y}
\right.\notag\\
&\hspace{30mm}\left.
+\frac{I_0(x)I_1(y)-I_1(x)I_0(y)}{x-y}
\right]
\notag\\
&=\frac{xy}{2}
\frac{xI_0(x)I_1(y)-yI_1(x)I_0(y)}{x^2-y^2},
\end{align}
which is Eq.~\eqref{eq:even-weighted-bessel-sum}.  Its apparent poles at
$x=\pm y$ are likewise removable, as is also clear from the convergent
series on the left-hand side.

\bibliographystyle{unsrt}
\bibliography{asymptotics}	

\begin{thebibliography}{10}

\bibitem{Berkooz:2020fvm}
Micha Berkooz, Nadav Brukner, Vladimir Narovlansky, and Amir Raz.
\newblock {Multi-trace correlators in the SYK model and non-geometric
  wormholes}.
\newblock {\em JHEP}, 21:196, 2020.

\bibitem{Sachdev:1992fk}
Subir Sachdev and Jinwu Ye.
\newblock {Gapless spin fluid ground state in a random, quantum Heisenberg
  magnet}.
\newblock {\em Phys. Rev. Lett.}, 70:3339, 1993.

\bibitem{Sachdev:2010um}
Subir Sachdev.
\newblock {Holographic metals and the fractionalized Fermi liquid}.
\newblock {\em Phys. Rev. Lett.}, 105:151602, 2010.

\bibitem{Kitaev2015SimpleModel}
Alexei Kitaev.
\newblock A simple model of quantum holography.
\newblock Talks at the KITP Program ``Entanglement in Strongly-Correlated
  Quantum Matter'', 2015.
\newblock Part 1, April 7; Part 2, May 27.

\bibitem{Maldacena:2016hyu}
Juan Maldacena and Douglas Stanford.
\newblock {Remarks on the Sachdev-Ye-Kitaev model}.
\newblock {\em Phys. Rev. D}, 94(10):106002, 2016.

\bibitem{Blommaert:2024ymv}
Andreas Blommaert, Thomas~G. Mertens, and Jacopo Papalini.
\newblock {The dilaton gravity hologram of double-scaled SYK}.
\newblock {\em JHEP}, 06:050, 2025.

\bibitem{Lin:2022rbf}
Henry~W. Lin.
\newblock {The bulk Hilbert space of double scaled SYK}.
\newblock {\em JHEP}, 11:060, 2022.

\bibitem{Eynard2004Topological}
Bertrand Eynard.
\newblock Topological expansion for the 1-hermitian matrix model correlation
  functions.
\newblock {\em Journal of High Energy Physics}, 2004(11):031, 2004.

\bibitem{EynardOrantin2007Invariants}
Bertrand Eynard and Nicolas Orantin.
\newblock Invariants of algebraic curves and topological expansion.
\newblock {\em Communications in Number Theory and Physics}, 1(2):347--452,
  2007.

\bibitem{EynardOrantin2009Topological}
Bertrand Eynard and Nicolas Orantin.
\newblock Topological recursion in enumerative geometry and random matrices.
\newblock {\em Journal of Physics A: Mathematical and Theoretical},
  42(29):293001, 2009.

\bibitem{Eynard2014Invariants}
Bertrand Eynard.
\newblock Invariants of spectral curves and intersection theory of moduli
  spaces of complex curves.
\newblock {\em Communications in Number Theory and Physics}, 8(3):541--588,
  2014.

\bibitem{EynardOrantin2007WeilPetersson}
Bertrand Eynard and Nicolas Orantin.
\newblock {Weil--Petersson} volume of moduli spaces, mirzakhani's recursion and
  matrix models, 2007.

\bibitem{Saad:2019lba}
Phil Saad, Stephen~H. Shenker, and Douglas Stanford.
\newblock {JT gravity as a matrix integral}.
\newblock 3 2019.

\bibitem{Okuyama:2025hsd}
Kazumi Okuyama.
\newblock {de Sitter JT gravity from double-scaled SYK}.
\newblock {\em JHEP}, 08:181, 2025.

\bibitem{Cotler:2016fpe}
Jordan~S. Cotler, Guy Gur-Ari, Masanori Hanada, Joseph Polchinski, Phil Saad,
  Stephen~H. Shenker, Douglas Stanford, Alexandre Streicher, and Masaki Tezuka.
\newblock {Black Holes and Random Matrices}.
\newblock {\em JHEP}, 05:118, 2017.
\newblock [Erratum: JHEP 09, 002 (2018)].

\bibitem{Khramtsov:2020bvs}
Mikhail Khramtsov and Elena Lanina.
\newblock {Spectral form factor in the double-scaled SYK model}.
\newblock {\em JHEP}, 03:031, 2021.

\bibitem{Saad:2018bqo}
Phil Saad, Stephen~H. Shenker, and Douglas Stanford.
\newblock {A semiclassical ramp in SYK and in gravity}.
\newblock 6 2018.

\bibitem{Liu:2025ikq}
Hong Liu.
\newblock {''Filtering'' CFTs at large N: Euclidean Wormholes, Closed
  Universes, and Black Hole Interiors}.
\newblock 12 2025.

\bibitem{Liu:2026rpw}
Hong Liu.
\newblock {Ramp, Plateau, and Wormholes without Averaging, and
  Hyper-non-perturbative Structures in Gravity}.
\newblock 8 2026.

\bibitem{Raz:2025wjw}
Amir Raz and Merna Youssef.
\newblock {The late time ramp from chord diagrams in the double-scaled SYK
  model}.
\newblock 7 2025.

\bibitem{Berkooz:2018qkz}
Micha Berkooz, Prithvi Narayan, and Joan Simon.
\newblock {Chord diagrams, exact correlators in spin glasses and black hole
  bulk reconstruction}.
\newblock {\em JHEP}, 08:192, 2018.

\bibitem{Berkooz:2018jqr}
Micha Berkooz, Mikhail Isachenkov, Vladimir Narovlansky, and Genis Torrents.
\newblock {Towards a full solution of the large N double-scaled SYK model}.
\newblock {\em JHEP}, 03:079, 2019.

\bibitem{Berkooz:2024lgq}
Micha Berkooz and Ohad Mamroud.
\newblock {A cordial introduction to double scaled SYK}.
\newblock {\em Rept. Prog. Phys.}, 88(3):036001, 2025.

\bibitem{Goel:2023svz}
Akash Goel, Vladimir Narovlansky, and Herman Verlinde.
\newblock {Semiclassical geometry in double-scaled SYK}.
\newblock {\em JHEP}, 11:093, 2023.

\end{thebibliography}
\end{document}